\documentclass[journal]{IEEEtran}
\usepackage{float}
\usepackage[caption=false]{subfig}
\usepackage[space]{grffile}
\usepackage{latexsym}
\usepackage{enumerate}
\usepackage{textcomp}
\usepackage{amsfonts,amsmath,amssymb}
\usepackage{bm}
\DeclareMathOperator*{\argmax}{\arg\!\max}
\usepackage{url}
\usepackage{algorithm}
\usepackage{algpseudocode}
\usepackage{tabularx}
\usepackage{longtable}
\usepackage{multirow,booktabs,makecell}
\usepackage{xcolor}
\usepackage{diagbox}
\usepackage{siunitx}
\usepackage[dvipdfm]{graphicx} \usepackage{bmpsize}
\usepackage[justification=centering]{caption}
\newcommand{\BO}{TPBO}

\newcommand{\fy}{y}

\newcommand{\fv}{\bm{\mathrm{f}}}
\newcommand{\fs}{\mathcal{S}}

\newcommand{\fa}{a(\bovar)}
\newcommand{\bovar}{\mathbf{q}}
\newcommand{\bovarspace}{\mathcal{Q}}

\newcommand{\Jnees}[0]{J_{NEES}}
\newcommand{\Jnis}[0]{J_{NIS}}

\def\mr[#1]#2#3{\multirowcell{#2}[#1]{#3}}

\newcommand{\Ignore}[1]{}

\newcommand{\kLst}{k-1}
\newcommand{\kCur}{k}

\newcommand{\x}[1]{\mathbf{x}_{#1}}
\newcommand{\z}[1]{\mathbf{z}_{#1}}
\newcommand{\m}[1]{\mathbf{m}_{#1}}

\newcommand{\pose}[1]{\mathbf{d}_{#1}}
\newcommand{\uVec}[1]{\mathbf{u}_{#1}}
\newcommand{\vVec}[1]{\mathbf{v}_{#1}}
\newcommand{\wVec}[1]{\mathbf{w}_{#1}}
\newcommand{\F}[1]{\mathbf{F}_{#1}}

\newcommand{\HM}[1]{\mathbf{H}_{#1}}
\newcommand{\HMt}[1]{\mathbf{H}_{#1}^\top}
\newcommand{\Q}[1]{\mathbf{Q}_{#1}}
\newcommand{\R}[1]{\mathbf{R}_{#1}}
\newcommand{\ex}[1]{\mathbf{e}_{\mathbf{x},#1}}
\newcommand{\ez}[1]{\mathbf{e}_{\mathbf{z},#1}}
\newcommand{\nees}[1]{\epsilon_{\mathbf{x},#1}}
\newcommand{\nis}[1]{\epsilon_{\mathbf{z},#1}}
\newcommand{\avgnees}[1]{\bar{\epsilon}_{\mathbf{x},#1}}
\newcommand{\avgnis}[1]{\bar{\epsilon}_{\mathbf{z},#1}}

\newcommand{\E}[1]{\mathrm{E}\left[#1\right]}

\newcommand{\ECondOuter}[2]{\E{{#1}{#1}^\top|#2}}
\newcommand{\xCond}[2]{\hat{\mathbf{x}}_{#1|#2}}

\newcommand{\covCond}[2]{\mathbf{P}_{#1|#2}}
\newcommand{\covPart}[2]{\mathbf{P}_{#1 #2}}

\newcommand{\zCond}[2]{\hat{\mathbf{z}}_{#1|#2}}

\newcommand{\innovCov}[2]{\mathbf{S}_{#1|#2}}
\newcommand{\nx}[0]{n_{\mathbf{x}}}
\newcommand{\nz}[0]{n_{\mathbf{z}}}
\newcommand{\Kw}[1]{\mathbf{K}_{#1}}

\newcommand{\Snu}[1]{\mathbf{S}_{#1}}

\ifCLASSINFOpdf

\else

\fi

\begin{document}

\title{Kalman Filter Tuning
with Bayesian Optimization}

\author{Zhaozhong~Chen,
        Nisar~Ahmed, 
        Simon~Julier, 
        and Christoffer~Heckman
        } 


\maketitle

\begin{abstract}

Many state estimation algorithms 
must be \emph{tuned:\/}
given the state space process and observation models, the process and
  observation noise parameters must be chosen. Conventional tuning approaches
  rely on heuristic hand-tuning or gradient-based optimization techniques to minimize a performance cost function.
  However, the relationship between tuned noise values and estimator
  performance is highly nonlinear and stochastic. Therefore, the tuning solutions can easily get trapped in local minima, which can lead to poor choices of noise parameters and suboptimal estimator performance.  This paper describes how Bayesian
  Optimization (BO) can overcome these issues. BO poses optimization as a
  Bayesian search problem for a stochastic ``black box'' cost function, where
  the goal is to search the solution space to maximize the probability of
  improving the current best solution. As such, BO offers a principled approach
  to optimization-based estimator tuning in the presence of local minima and
  performance stochasticity.  While extended Kalman filters (EKFs) are the main
  focus of this work, BO can be similarly used to tune other related state
  space filters.  The method presented here uses performance metrics derived
  from normalized innovation squared (NIS) filter residuals obtained via
  sensor data, which renders knowledge of ground-truth states unnecessary. The
  robustness, accuracy, and reliability of BO-based tuning is illustrated on
  practical nonlinear state estimation problems, losed-loop aero-robotic control. 

\end{abstract}


\IEEEpeerreviewmaketitle

\section{Introduction}


\IEEEPARstart{M}{any} state estimation algorithms, including Kalman filters and particle filters, are recursive and model-based \cite{wan2000unscented}, \cite{li2001practical}. They decompose the estimation problem into a cycle with two main steps: \emph{state prediction} followed by \emph{measurement update}. The state prediction step uses a process model to predict how the state evolves over time. The measurement update step uses an observation model to relate a measured quantity to the state estimate. Since both the process and observation models are imperfect, errors in these models are treated as random noise terms that are injected into the system. Most designs assume the noises are white, zero mean and uncorrelated. As a result, filter tuning consists of choosing the values of the process and observation noise covariances.

Many tuning procedures adopt a divide-and-conquer strategy. The first stage is to choose the observation covariance. This is normally carried using laboratory or bench testing. The sensor is placed in a condition in which the noise-free sensor values can be predicted. The observation noises are determined by statistically characterizing the difference between the predicted and actual values. In the second stage, the process noises are chosen. Since the process noises contain information about the state disturbances and dynamic model uncertainties , which often cannot be reproduced in laboratory settings, the covariance is often chosen by collecting data from an operational domain and quantifying the quality of the estimates. Typically a performance cost is assigned, and the process noise covariance adjusted to minimize the value of that cost.

However, there are several problems with this two-stage approach. First, with laboratory testing, it is not always possible to model how sensors will react in operational environments. Changes in temperature, for example, can cause the biases in IMUs to change. As a result, the observation noises might not be properly characterized. Second, the interaction between noise levels and filter performance is not straightforward. Theoretical analysis has shown that even if the process and observation models are linear, the presence of modeling errors lead to noises which are state-dependent and correlated over time \cite{dette2016optimal, wanninger1998real}. As a result, many non-unique tuning solutions can appear. Finally, they tend to rely on statistics which require knowledge of the ground truth of the
system to compute.  Although this is possible to obtain in simulations or
laboratory settings with precise reference measurement systems, such approaches
are of limited use for many practical real-world applications where ground
truth data is not available. 

This paper makes three contributions, which significantly extends the
preliminary work presented in \cite{chen2018weak}. The first is a general
framework for Kalman filter noise parameter tuning based on Bayesian
Optimization (BO).  BO provides an attractive way to
overcome the limitations of other gradient-based optimal filters auto-tuning
strategies, which can easily get trapped in poor local minima.  
The BO framework developed here uses nonparametric surrogate models based on
Student-t process regression, which offers better robustness and performance
compared to Gaussian process surrogate models that are more typically used for
Bayesian optimization and which were considered in \cite{chen2018weak}.  

The second contribution is the development and validation of novel stochastic cost functions for optimization-based auto-tuning. Most auto-tuning algorithms including our previous work \cite{chen2018weak} use the Normalized Estimation Error Squared (NEES) which requires the ground truth values of the state. We may not be able to obtain the ground truth easily in the real world. To make the tuning process be more easily to implement in the real world, in this paper, our approach extends the previous work and uses Normalized Innovation Squared (NIS),  which is computed from the difference between the predicted and actual sensor measurements and does not require ground truth.
This makes our approach practical for many real-world
applications, where only the sensor observations are available for filter
validation. 

Finally, the performance of the BO auto-tuning
framework is demonstrated and evaluated in simulation for a
challenging application: longitudinal state estimation for 
the Mars Science Laboratory (MSL) Skycrane platform. The results for this
application shows that BO not only provides reliable and
computationally efficient estimates of unknown filter parameters, but can also
provide useful probabilistic information about each parameter through the whole
domain space, which existing state-of-the-art auto-tuning methods cannot do. While this work focuses mainly on auto-tuning of extended Kalman filters (EKFs), the BO framework can be readily extended to other related state space filtering algorithms.

The rest of this paper is structured as follows. 
Sections~\ref{sct:problem_statement} and \ref{sct:tuning} formally introduces an overview and problem statement for filter auto-tuning. Section~\ref{sct:bayesopt} describes our Bayesian optimization framework using nonparametric Student-t process regression and how it is applied to Kalman filter parameter auto-tuning. Section~\ref{sct:result} describes the set up, and analysis for the numerical simulation studies using the Bayesian optimization framework to auto-tune EKFs for the Mars Skycrane longitudinal state estimation problem. 
Conclusions are given in Section \ref{sct:final}. 

\section{Preliminaries}
\label{sct:problem_statement}

\subsection{System Overview}

The state of the system at time step $k$ is $\x{k} \in \mathbb{R}^{n_\mathbf{x}}$, where $n_\mathbf{x}$ is the dimension of the state vector. The process model that propagates the state from $\kLst$ to $\kCur$ is
\begin{equation}\label{eq:dynModel}
    \x{\kCur} = f(\x{k-1}, \uVec{k}, \vVec{k})
\end{equation}
where $\uVec{k} \in \mathbb{R}^{n_\mathbf{u}}$ is the control input vector and $\vVec{k} \in \mathbb{R}^{n_\mathbf{v}}$ is the process noise. 

The observation model is
\begin{equation}\label{eq:measModel}
    \z{k} = h(\x{k}, \wVec{k})
\end{equation}
\noindent where $\z{k} \in \mathbb{R}^{n_\mathbf{z}}$ is the observation vector and $\mathbf{w}_{k} \in \mathbb{R}^{n_\mathbf{z}}$ is the measurement noise.

\Ignore{$\F{k}$ is the state transition matrix, which
comes from the Jacobian of the process model with respect to the state.
$\HM{\kCur}$ is the observation transition matrix, which comes from the
Jacobian of observation model with respect to the state. $\F{k}$ and
$\HM{\kCur}$ may be written as:

\begin{align}
    {{\F{k}}} &= \left . \frac{\partial f}{\partial \mathbf{x} } \right \vert _{\xCond{k-1}{k-1},\uVec{k}}, \label{process_jacobian}\\
    {\HM{k}} &= \left . \frac{\partial h}{\partial \mathbf{x} } \right \vert _{\xCond{k}{k-1}}, \label{measurement_jacobian}
\end{align}}

It is typically assumed that the process and observation models are sufficiently accurate that the process and observation noises are zero-mean, independent, Gaussian distributed random variables.

Given the structure of the system, the goal is to develop an estimation algorithm which takes in a sequence of observations and control inputs, and computes an estimate of the state. The errors in initial conditions, together with the process and observation noises, means that the state is not known perfectly. Therefore, some means of quantifying the uncertainty must be used. A common approach is to use the mean and covariance of the state estimate.

\subsection{Mean and Covariance Representation} \label{tuning_basic}

Our goal is to estimate the state of a random variable $\x{i}$ at a discrete
time $i$ and quantify the uncertainty $\covCond{i}{i}$ in that estimate. Let
$\xCond{i}{j}$ be the estimate of $\x{i}$ using all observations up to time step $j$, and the covariance of this estimate be $\covCond{i}{j}$:\\
\begin{align}
\xCond{i}{j}&=\E{\x{i}|\z{1:j}}\\
\covCond{i}{j}&=
\ECondOuter{\left(\x{i}-\xCond{i}{j}\right)}{\z{1:j}}.
\end{align}

However, computing an estimate which obeys this property in practice is difficult to achieve. Modelling errors, for example, can always lead to biased estimates. Therefore, most pratical systems use a weaker condition called \emph{covariance consistency} \cite{uhlmann2003covariance}. In this case, a valid estimate has the properties:
\begin{align}
\xCond{i}{j}&\approx\E{\x{i}|\z{1:j}}\\
\covCond{i}{j}&\ge
\ECondOuter{\left(\x{i}-\xCond{i}{j}\right)}{\z{1:j}}.
\end{align}
\noindent where $\approx$ is application-specific and $\mathbf{A}\ge\mathbf{B}$ means
that $\mathbf{A}-\mathbf{B}$ is positive semidefinite. In other words, the
estimate should be approximately unbiased, and the estimator should not over
estimate its level of confidence. At the same time, we would like the difference between the predicted covariance and actual mean squared error to be as small as possible.

\subsection{Kalman Filters}


The Kalman filter is one of the best known and most widely used algorithms for state estimation. It is derived from the fact that the correction applied to the estimate is a linear  rule of the form
\begin{equation*}
    \xCond{\kCur}{\kCur}=\xCond{\kCur}{\kLst}+\Kw{\kCur}\ez{\kCur},
\end{equation*}
where $\Kw{\kCur}$ is a gain matrix, and
\begin{equation*}
    \ez{\kCur}=\z{\kCur}-\zCond{\kCur}{\kLst},
\end{equation*}
which is the difference between the actual and predicted sensor measurement, is the innovation vector. It acts as an error signal in the filter, and provides a correction term for the state estimate. The Kalman filter chooses the value of $\Kw{\kCur}$ to minimize the mean squared error in $\xCond{\kCur}{\kCur}$.

The algorithm proceeds as follows~\cite{Bar-Shalom2001}. The state is predicted according to the equation
\begin{align}
    \xCond{k}{k-1} &= f(\xCond{k-1}{k-1}, \uVec{k})\\
    \covCond{k}{k-1} &=  {\F{k}} \covCond{k-1}{k-1}{ {\F{k}^\top}} + \Q{k}.
\end{align}
The update is calculated from
\begin{align}
    \xCond{\kCur}{\kCur}&=\xCond{\kCur}{\kLst}+\Kw{\kCur}\ez{\kCur}, \label{update1_dis1}\\
    \covCond{\kCur}{\kCur}&= (\mathbf{I} -\Kw{\kCur}\HM{\kCur})\covCond{k}{k-1}, \label{update1_dis2}\\
    \Snu{\kCur|\kLst}&=\HM{\kCur}\covCond{\kCur}{\kLst}\HMt{\kCur}+\R{\kCur}, \label{update1_dis3}\\
    \Kw{\kCur}&=\covCond{\kCur}{\kLst}\HMt{\kCur}\Snu{\kCur|\kLst}^{-1},  \label{update1_dis4}
\end{align}
where $\F{k}$ and $\HM{k}$ are the Jacobian matrices of the process and observation models.

\begin{align}
    {{\F{k}}} &= \left . \frac{\partial f}{\partial \mathbf{x} } \right \vert _{\xCond{k-1}{k-1},\uVec{k}} \label{process_jacobian}\\
    {\HM{k}} &= \left . \frac{\partial h}{\partial \mathbf{x} } \right \vert _{\xCond{k}{k-1}} \label{measurement_jacobian}
\end{align}

Note that once Eq.\ \eqref{eq:dynModel} and \eqref{eq:measModel} have been chosen,  the only degree of freedom left is to chose $\Q{k}$ and $\R{\kCur}$. This process is known as tuning.


\section{Tuning}
\label{sct:tuning}

As explained in the introduction, tuning involves choosing $\Q{k}$ and $\R{k}$ to minimize some performance cost. Two widely used measures are the \emph{normalized estimation error squared (NEES)} and the \emph{normalized innovation error squared (NIS)}. These are computed from
\begin{align}
\nees{k} &= \ex{k}^T \covCond{k}{k}^{-1} \ex{k} \label{eq:neesDef}   \\
\nis{k} &= \ez{k}^T \innovCov{k}{k-1}^{-1} \ez{k} \label{eq:nisDef}
\end{align}
where $\ex{k} = \xCond{k}{k} - \x{k}$ and $\ez{k} =
\mathbf{z}_{k} - h(\hat{\mathbf{x}}_{k|k-1})$ is the innovation vector. If the dynamical consistency conditions are met, then 
\begin{align}
    \E{\nees{k}}{\approx}\nx\\
    \E{\nis{k}}{\approx}\nz.
\end{align}
It is often assumed that the prediction and observation errors are Gaussian. In this case, $\nees{k}$ and $\nis{k}$ will be $\chi^2$-distributed random variables with
$\nx$ and $\nz$ degrees of freedom respectively \cite{Bar-Shalom2001}. Therefore, $\chi^2$ hypothesis tests can be performed on calculated values for
$\nees{k}$ (when ground truth data is available) and $\nis{k}$ to see if the
consistency conditions hold at each time $k$.

\subsection{Approaches to Tuning}


In this paper we focus on the process noise tuning because it is the hardest to tune in the Kalman filter. In practice, NEES $\chi^2$ tests are conducted using multiple offline Monte
Carlo ``truth model'' simulations to obtain ground truth $\x{k}$ values. The
truth model simulator represents a high-fidelity model of the ``actual'' system
dynamics and sensor observations, which may contain nonlinearities and other
non-ideal characteristics that must be compensated for via Kalman filter
tuning. NIS $\chi^2$ can be conducted offline using multiple Monte Carlo
simulations (e.g.\ in parallel with NEES tests), but can also be conducted
online using real-time sensor data.

Online/offline NIS tests are conducted as follows \footnote{This is the same as
the offline truth model tests conducted in \cite{chen2018weak}; here we
still use ground truth in order to check if the filter is consistent but in
practice it is not required.}: suppose $N$ independent instances of the
true state are randomly initialized according to $\xCond{0}{0}$ and
$\covCond{0}{0}$ (the initial state of the filter), and then propagated through
the true stochastic dynamics \eqref{eq:dynModel} and measurement model
\eqref{eq:measModel} for $T$ time steps, yielding sample ground truth sequences
$\x{1}^i,\x{2}^i,\ldots,\x{T}^i$ and measurement sequences
$\z{1}^i,\z{2}^i,\ldots,\z{T}^i$ for $i=1,\ldots,N$. If the resulting
measurement sequences are then fed into a Kalman filter with tuning parameters
$(\Q{k},\R{k})$, the resulting NEES and NIS statistics for each simulation run
$i$ at each time $k$ can be averaged across problem instances to give the test
statistics:

\begin{align}
\avgnees{k} &= \frac{1}{N} \sum_{i=1}^{N}{\nees{k}^i} \label{eq:neesAvgk} \\
\avgnis{k} &= \frac{1}{N} \sum_{i=1}^{N}{\nis{k}^i}. \label{eq:nisAvgk}
\end{align}

Then, given some desired Type I error rate $\alpha$, the NEES and NIS $\chi^2$
tests provide lower and upper tail bounds
$[l_{\mathbf{x}}(\alpha,N),u_{\mathbf{x}}(\alpha,N)]$ and
$[l_{\mathbf{z}}(\alpha,N),u_{\mathbf{z}}(\alpha,N)]$, such that the Kalman
filter tuning is declared to be consistent if, with probability $100(1-\alpha)$
at each time $k$,

\begin{align*}
&\avgnees{k} \in [l_{\mathbf{x}}(\alpha,N),u_{\mathbf{x}}(\alpha,N)],  \\
&\avgnis{k} \in [l_{\mathbf{z}}(\alpha,N),u_{\mathbf{z}}(\alpha,N)].
\end{align*}

Otherwise, the filter is declared to be inconsistent.  Specifically, if
$\avgnees{k}<l_{\mathbf{x}}(\alpha,N)$ or
$\avgnis{k}<l_{\mathbf{z}}(\alpha,N)$, then the filter tuning is ``pessimistic''
(underconfident), since the filter-estimated state error/innovation
covariances are too large relative to the true values.  On the other hand, if
$\avgnees{k}>u_{\mathbf{x}}(\alpha,N)$ or
$\avgnis{k}>u_{\mathbf{z}}(\alpha,N)$, then the filter tuning is ``optimistic''
(overconfident), since the filter-estimated state error/innovation covariance
are too small relative to the true values.

The $\chi^2$ consistency tests provide a very principled basis for validating Kalman filter performance in domain-agnostic way, and also provide a well-established means for guiding the tuning of noise parameters $\Q{k}$ and $\R{k}$ in practical applications. Tuning via the $\chi^2$ tests is most often done manually, and thus requires repeated ``guessing and checking'' over multiple Monte Carlo simulation runs. However, this quickly becomes cumbersome and non-trivial for systems with several tunable noise terms. Heuristics for manual filter tuning have been developed in the linear-quadratic optimal control literature \cite{stengel1986optimal}, e.g. to coarsely tune diagonals of $\Q{k}$ first, before fine-tuning the elements of $\Q{k}$ further. Such heuristics are useful for bounding the shape and magnitude of $\Q{k}$ in linear-Gaussian problems, but are of little help for tuning `fudge factor' process noise parameters that are used to cope with model errors from state truncation, approximations of non-linearities, poorly modeled dynamics, etc. 
Given this, 
alternative optimization techniques are needed which are robust to stochastic variations in the cost function and which can explore nonlinear spaces while also satisfying the filter consistency requirements. 


\subsection{Previous Tuning Work} \label{previous_work}


Much of the previous Kalman filter auto-tuning work is based on consistency checking (\cite{oshman2000optimal} and \cite{powell2002automated}).\\
\indent Reference \cite{oshman2000optimal} uses a genetic algorithm to tune Kalman filter. This algorithm simulates the Darwin concept of ``survival of the fittest'' to choose a good parameter set. It treats each parameter set in the parameter space as an ``individual.'' The specific parameter value corresponding to that individual is coded into a string as a binary value and treated as a ``chromosome,'' which is the genetic information. The fittest individual is selected  according to the numerical value of NEES and error covariance norm, which is the cost function.   The genetic algorithm is implemented after some modifications. First, random parameter sets in the parameter space are selected as the initial ``population.'' They will spawn the next generation by pairing two individuals and exchanging parts of their chromosomes randomly.
The population is believed to have converged once the population has a low cost. In this approach,  a large number of Monte Carlo runs is not used because of computation limits, which leads to a problem that some wrong individuals may also be able to pass the consistency test. They add one more option to the cost function besides the consistency test to solve that problem: when the consistency value is smaller than a threshold the cost function switches to a value based on the norm of the error covariance $\mathbf{P}$.\\
\indent In their simulation experiment, they use a simple oscillator as an example, aiming at tuning the speed and position noise. The optimal value is not achieved because the structure of their cost function: the minimization routine will tend to have smaller state error covariance norm and instead of smaller consistency value.\\
\indent Another previous work \cite{powell2002automated} uses downhill simplex numerical optimization algorithm to minimize the NEES based cost function. A simplex is a collection of $N+1$ points in an $N$-dimensional space and all their interconnecting line segments. The simplex algorithm attempts to locate a minimum of the function by a series of movements in the $N$-dimensional space. Those movements include reflection, expansion and contraction. Details of those movements can be seen in the paper \cite{powell2002automated}.\\ 
\indent However, the simplex algorithm can easily be stuck in a local minima so there may be cases that this method will fail. Although the algorithm's cost function is based on NEES, there are no plots showing the consistency check after getting a tuning result.\\ 
\indent Our previous work \cite{chen2018weak} focus on Kalman filter tuning using NEES $\chi^2$ tests too. Due to the hardware improvements these years, it is not that time consuming to implement a large number of Monte Carlo tests consisting of, say, several hundred runs. We simulated a car moving along a straight line and optimized a two dimensional process noise and a one dimensional measurement noise. We successfully showed that use Bayesian optimization to tune Kalman filter process noise covariance as well as measurement noise covariance and can yield good results. However, in our previous work, we did not perform formal post hoc consistency validation  checks to confirm the readers that the error at each timestamp is small enough. Our previous work also only limited analysis and application to a linear dynamical system, and did not consider extensions to linearization-based approximations for non-linear filtering. At the same time, the above mentioned references and our prior work \cite{chen2018weak} use NEES based consistency check method, which makes it impossible to use when the ground truth is not available. In this paper, we also propose to use a NIS based consistency check method. NIS based tuning method makes it possible for us to tune the Kalman filter with just sensor measurements. We apply our methods on more complicated and practical nonlinear cases and also validate statistical consistency of the optimized result. Finally, in this paper we use an improved Bayesian optimization procedure which is based on nonparametric Student's-t regression models, which leads to significantly more robust surrogate models and tuning solutions than the Gaussian processes (GPs) regression models used in our prior work. 

\subsection{Summary}

Problems with existing approaches are that (a) they fall into local minima; (b) they often have to use NEES; (c) they run into issues with noise and stochastic variation from small finite number of MC runs. We use Bayesian optimization to avoid falling into local minima and we use NIS to avoid using groundtruth.

\section{Bayesian Optimization for auto tuning}
\label{sct:bayesopt}


Many approaches for solving nonlinear optimization problems use gradient descent. However, the risk with these approaches is that they can fall into local minima. This issue is exacerbated for filter tuning problems defined by noisy nonlinear dynamical systems. 
Stochastic variations and nonlinear model characteristics can introduce many local minima into objective functions for tuning that can trap gradient descent methods. One principled way to handle such cases is to use \emph{Bayesian optimization\/} \cite{pelikan1999boa}, which poses optimization as a probabilistic search problem.

Bayesian optimization is first described for dealing with generic ``black box'' stochastic objective functions. Its novel application is then described for simulation-based Kalman filter auto-tuning. 
\subsection{Bayesian Optimization Theory}
Consider the minimization of an objective function $y:\bovarspace \rightarrow \mathbb{R}$, where $\bovarspace \in \mathbb{R}^d$ is the search or solution space, and $\bovar^* \in \bovarspace$ is the minimizer, such that $y(\bovarspace^*) \leq y(\bovarspace), \  \forall \bovar \in \bovarspace$.
Furthermore, we assume that each elements of $i$ of $\bovar$ lies in the interval $\bovarspace (i) \in [\bovar(i)_l, \bovar(i)_u]$.

The intuition behind BO arises from the following. First suppose that the entire solution space were densely sampled. Carrying out this process, the map $y:\bovarspace \rightarrow \mathbb{R}$ is entirely known and the minimizer can be read off directly. However, this dense sampling scheme is not possible in practice. Rather, the search algorithm samples a subset of the parameter space. Since the sampling is incomplete, the shape of the cost surface is not known but, rather, must be estimated from sparse and incomplete data.
Therefore, the goal of Bayesian optimization is to find the minimizer of the noisy objective function $\fy$ 
while at the same time learning about the mapping from $\bovar$ to $\fy$ via Bayesian inference. Bayesian optimization uses ``black box'' point evaluations of $y$ to efficiently find $\bovar^*$. This is accomplished by maintaining beliefs about how $\fy$ behaves over all $\bovar$ in the form of a surrogate model $\fs$, which statistically approximates $\fy$ and is easier to evaluate (e.g. since evaluations of $\fy$ might require expensive high-fidelity simulation). During optimization, $\fs$ is used to determine where the next design point sample evaluation of $\fy$ should occur, in order to update beliefs over $\fy$ and thus simultaneously improve $\fs$, while finding the (expected) minimum of $\fy$ as quickly as possible. The key idea is that, as more observations $E$ are sampled at different $\bovar$ locations, the $\bovar$ samples themselves eventually converge to the expected minimizer $\bovar^*$ of $\fy$. Since $\fs$ contains statistical information about the level of uncertainty in $\fy$ (i.e. related to the posterior belief $p(y|E)$), Bayesian optimization effectively leverages probabilistic ``explore-exploit'' behavior to learn a probabilistic model of $\fy$ while also minimizing it.

We next describe the two main components of the Bayesian optimization process: (1) the surrogate model $\fs$, which encodes statistical beliefs about $\fy$; and (2) the acquisition function $\fa$, which is used to intelligently guide the search for $\bovar^*$ via $\fs$. 

\subsubsection{Surrogate Model}

The surrogate model is the model used to approximate the objective function. In the BO literature, nonparametric regression models based on stochastic processes are widely used~\cite{rasmussen2003gaussian} because they naturally model probability distributions over uncertain functions and can be evaluated at arbitrary query points given some finite set of sample observation points. 
Although Gaussian process (GP) are frequently used, in this work we use Student-t processes (TP) instead, following the recommendation of Shah, et al.\ \cite{shah2014student}. 

\indent A Student-t process is a stochastic process such that every finite collection of samples from the process has a multivariate Student-t joint distribution. The mean function $\Phi(\mathbf{q})$, the kernel function $k(\mathbf{q})$ and the parameter $v$ are the main characteristics of TP. It can be written as 
\begin{equation}
    y(\mathbf{q}) \sim \mathcal{TP}(v, \Phi(\mathbf{q}), k(\mathbf{q},\mathbf{q}')).
\end{equation}
The real-valued parameter $v>2$ controls how ``heavy-tailed'' the process is. The heavier the tail (i.e. the smaller the $v$), the more likely it is that the TP will produce a value that is far from the mean value. The TP tends to a GP as $v \rightarrow \infty$. 
The TP is attractive because it provides some extra benefits over GP, without incurring more computational cost. For example, the predictive covariance for the TP explicitly depends on observed $y$ data values; this is a useful property which the Gaussian process lacks. Furthermore,  distributions over the cost function $y$ may in general be heavy-tailed, so it is better to use TP to ``safely'' model their behaviors \cite{shah2013bayesian}. Similarly, every finite collection of TP samples $\mathbf{q}_{1:n} = (\mathbf{q}_1, \mathbf{q}_2, \cdots, \mathbf{q}_n)$ has a multivariate Student-t distribution, which can be written as \\
\begin{align} \label{mvt}
\begin{split}
    & y(\mathbf{q}_{1:n}) = \frac{\Gamma(\frac{v+n}{2}) }{\Gamma(v/2)((v-2)\pi)^{n/2}|K|^{1/2}} \\
    & \times (1 + \frac{1}{v-2}(\mathbf{q}_{1:n} - \Phi(\mathbf{q}_{1:n})^TK^{-1}(\mathbf{q}_{1:n} - \Phi(\mathbf{q}_{1:n}))^{-\frac{v+n}{2}}
\end{split}
\end{align}
\noindent where $\Gamma(n)$ is the gamma function for $n\in \mathbb{R}$, 
and $v>2$. $K$ is the covariance matrix consisting of kernel function $k$ evaluations, 
\begin{equation} \label{cov_matrix}
    K = 
    \begin{bmatrix}
    k(\mathbf{q}_1, \mathbf{q}_1) & k(\mathbf{q}_1, \mathbf{q}_2) & \cdots & k(\mathbf{q}_1, \mathbf{q}_n) \\
    k(\mathbf{q}_2, \mathbf{q}_1) & k(\mathbf{q}_2, \mathbf{q}_2) & \cdots & k(\mathbf{q}_2, \mathbf{q}_n) \\
    \vdots & \vdots & \vdots & \vdots \\
    k(\mathbf{q}_n, \mathbf{q}_1) & k(\mathbf{q}_n, \mathbf{q}_2) & \cdots & k(\mathbf{q}_n, \mathbf{q}_n)
    \end{bmatrix}.
\end{equation}
Equation (\ref{mvt}) is written as the following for simplicity,
\begin{equation}
    y(\mathbf{q}_{1:n}) \sim MVT_n(v, \Phi(\mathbf{q}_{1:n}), K).
    \label{eq:mvt}
\end{equation}
For Bayesian optimization, newly sampled $\bovar$ and $y$ values are added to the vector $\mathbf{q}_{1:n}$ and $y(\mathbf{q}_{1:n})$ to construct a surrogate model of the underlying objective function according to Eq.\ (\ref{eq:mvt}). In most implementations of Bayesian optimization, as the new sample values are added, the hyperparameters for the kernel function $k$ are also re-estimated from the available data and updated accordingly. The updated surrogate model is then used to compute the acquisition function, which is used to select the next sample $\bovar$ for evaluation of $y$. 

\subsubsection{Acquisition Function}
The expected improvement function is one of many well-known acquisition functions; other possible and popular choices for the acquisition function include the \emph{Lower Confidence Bound (LCB)}. We use expected improvement function in of our implementations and we'll discuss how this acquisition function is generated next. Suppose that $n$ points $\mathbf{q}_{1:n}$ and $y(\mathbf{q}_{1:n})$ have been sampled and have been incorporated into the surrogate model. The current minimizer is $\mathbf{q}^*_{1:n}=\min_{m \leq n}y(\mathbf{q}_m)$. This will be one of these points, since observations of $y$ are not available for other points in $\bovarspace$. The algorithm needs to choose the next point $\mathbf{q}_{n+1}$ to be sampled. We seek a new location which will yield a new, lower, minimum. In other words, $y(\mathbf{q}_{n+1})<y(\mathbf{q}^*_{1:n})$. 
One way to evaluate the new sample point is to evaluate its 
\emph{improvement} with respect to $\mathbf{q}^*_{1:n}$
\begin{equation}
    g(\mathbf{q}_{n+1}, \mathbf{q}^*_{1:n}) = \max(0, y_n(\mathbf{q}^*_{1:n}) - y(\mathbf{q}_{n+1})).
\end{equation}
This only assigns a non-zero value if  $\mathbf{q}_{n+1}<\mathbf{q}^*_{1:n}$. Therefore, the idea is to chose $\mathbf{q}_{n+1}$ which maximizes the improvement. Since we only have access to the surrogate function, the improvement is stochastic. Therefore, we choose the next sample point based on the \emph{expected improvement} 
\begin{equation} \label{expected_improvement}
    E_n[g(\mathbf{q}_{n+1}, \mathbf{q}^*_{1:n}) \ | \ \mathbf{q}_{1:n}, \mathbf{y}(\mathbf{q}_{1:n}) ]
\end{equation}
\noindent $E_n[\cdot]$ is the expectation based on current posterior distribution, given by the current $MVT$ surrogate model. We need maximize Eq.\ \ref{expected_improvement} to find the next point to sample.
\begin{equation}
    \mathbf{q}_{n+1} = \argmax\limits_{\mathbf{q}_{n+1}} E_n[g(\mathbf{q}_{n+1}, \mathbf{q}^*_{1:n}) \ | \ \mathbf{q}_{1:n}, \mathbf{y}(\mathbf{q}_{1:n}) ]
\end{equation}
The closed form solution of the $\mathrm{EI}$ acquisition function using the TP surrogate is \cite{shah2013bayesian, tracey2018upgrading},
\begin{align} \label{EI_function}
    \mathrm{EI}_n(\mathbf{q}) = (y_n(\mathbf{q}^*_{1:n}) - u)\Psi(z) + \frac{v}{v-1}(1 + \frac{z^2}{v})\sigma \psi(z)
\end{align}
\noindent where $u$ and $\sigma$ are the mean and variance of the conditional Student's-t distribution of $\mathbf{q}_{n+1}$, which is presented below in \eqref{con_MVT}.  $\Psi(\cdot)$ and $\psi(\cdot)$ are the CDF and PDF of the standard Student-t distribution $MVT_1(v, 0, 1)$. The conditional $MVT$ distribution is similar to the conditional multivariate Gaussian distribution: if we have $\mathbf{q}_{1:n+1}$ and $y(\mathbf{q})_{1:n+1}$ described by a multivariate $MVT$ pdf, then 
\begin{align} \label{MVT_nplus1}
    \begin{split}
    \begin{bmatrix}
        y(\mathbf{q}_{1:n}) \\
        y(\mathbf{q}_{n+1})
    \end{bmatrix}
    & \sim 
    MVT_{n+1} (v+1,
    \begin{bmatrix}
        \Phi(\mathbf{q}_{1:n}) \\
        \Phi(\mathbf{q}_{n+1}) \\
    \end{bmatrix} 
    , \\
    & \begin{bmatrix}
        K(\mathbf{q}_{1:n}, \mathbf{q}_{1:n}) & K(\mathbf{q}_{1:n}, \mathbf{q}_{n+1}) \\
        K(\mathbf{q}_{n+1}, \mathbf{q}_{1:n}) & K(\mathbf{q}_{n+1}, \mathbf{q}_{n+1}) \\
    \end{bmatrix} )
    \end{split}
\end{align}
where $K(\mathbf{q}_{1:n}, \mathbf{q}_{1:n})$ is the same as eq. (\ref{cov_matrix}), $K(\mathbf{q}_{1:n}, \mathbf{q}_{n+1}) = [k(\mathbf{q}_1 , \mathbf{q}_{n+1}), \cdots, k(\mathbf{q}_n , \mathbf{q}_{n+1})]^T$, and $K(\mathbf{q}_{n+1}, \mathbf{q}_{n+1}) = k(\mathbf{q}_{n+1} , \mathbf{q}_{n+1})$. 
\eqref{MVT_nplus1} can be written more simply as
\begin{equation} \label{sim_MVT_nplus1}
    \begin{split}
    \begin{bmatrix}
        y_1 \\
        y_2
    \end{bmatrix}
    \sim 
    MVT_{n+1}(v+1,
    \begin{bmatrix}
        \Phi_1 \\
        \Phi_2 \\
    \end{bmatrix}
    , 
    \begin{bmatrix}
        K_{11} & K_{12} \\
        K_{21} & K_{22} \\
    \end{bmatrix} )
    \end{split}
\end{equation}
The conditional Student-t distribution of $y(\mathbf{q}_{n+1})$ is then given by \cite{ding2016conditional}
\begin{equation} \label{con_MVT}
    \begin{split}
        & y(\mathbf{q}_{n+1}|\mathbf{q}_{1:n}, y(\mathbf{q}_{1:n})) \sim MVT_1(v+n, u, \sigma) \\
        & u = \Phi_2 + K_{21}K_{11}^{-1}(y_1 - \Phi_1) \\
        & \sigma = \frac{v+d}{v+n} K_{22}-K_{21}K_{11}^{-1}K_{12} \\
        & d = (y_1 -  \Phi_1)^T K_{11}^{-1}(y_1 - \Phi_1)
    \end{split}
\end{equation}
The prior mean function for the surrogate model in Bayesian optimization can be set as a constant 
without changing the final result \cite{brochu2010tutorial}, 
so let $\Phi_1 = \mathbf{0}, \Phi_2 = 0$. Substituting Eq.\ (\ref{con_MVT}) into Eq.\ (\ref{EI_function}), the only unknown variable will be $\mathbf{q}_{n+1}$. To find the $\mathbf{q}_{n+1}$ that maximizes Eq.\ \eqref{EI_function} for the largest improvement, another inner optimization problem must be solved within Bayesian optimization. Luckily, Eq.\ \eqref{EI_function} is known, there are several ways to maximize the $\mathrm{EI}$ function, such as \emph{DIRECT} \cite{finkel2003direct}, which is a derivative free and deterministic nonlinear global optimization algorithm that is widely used for Bayesian optimization via nonparametric surrogate model regression. Once the next point to sample is selected by the inner loop optimization, the Bayesian optimization loop can continue until the termination criterion (maximum iteration or minimum observation change between two iterations) is met.

The resulting algorithm for Bayesian optimization is referred to here as \BO{} (Student-t processes Bayesian optimization). 
 $\mathrm{EI}$ is used in this work 
because it has been shown to yield better or equal performance to other acquisition functions for a wide variety of applications \cite{gardner2014bayesian}, \cite{gelbart2014bayesian}, \cite{wang2013bayesian}. 

\subsection{Stochastic Costs for Consistency-based Filter Auto-tuning} \label{sec:stochcosts} 
Now consider how $\fy(\bovar)$ can be defined via NEES and NIS consistency test statistics for Kalman filter tuning. As such, let $\bovarspace$ be some space of configurable Kalman filter parameters (e.g. the set of all parameters defining some positive definite symmetric process noise covariance $\Q{k}$) and let $\bovar \in \bovarspace$ be a design point. 

Consider first the case of tuning based on assessment of NEES statistics obtained via Monte Carlo ground truth simulation models. If $N$ Monte Carlo simulations are performed for $T$ time steps at any given design point $\bovar$, starting from the initial conditions $\xCond{0}{0}$ and $\covCond{0}{0}$, then the average NEES statistic $\avgnees{k}$ can be computed using Eq.\  \eqref{eq:neesAvgk} for each time $k=1,...,T$. 
To summarize how ``well-behaved'' $\avgnees{k}$ is across all time steps, we can leverage the fact that the expected value of $\avgnees{k}$ for a consistent Kalman filter should be $\nx$. 
Therefore, we use the following cost function to evaluate how much $\avgnees{k}$ deviates from this ideal expected value
\begin{align}
\fy(\bovar) = \Jnees(\bovar) =  \left|  \log \left(\frac{\sum_{k=1}^{T}{\avgnees{k}}/T}{\nx} \right)  \right|
\end{align}

The reason why we use the log of the cost is that the NEES value itself is bounded from below (by 0) but is not bounded above. Taking the log ensures that the cost will space from negative to positive infinity.

By a similar reasoning,  
\begin{align}\label{JNIS_def}
\fy(\bovar) = \Jnis(\bovar) =  \left| \log \left(\frac{\sum_{k=1}^{T}{\avgnis{k}}/T}{\nz} \right)  \right|
\end{align}
where $\avgnis{k}$ is the NIS outcome which could either be obtained from a ground truth simulation, or from a set of real data logs. One can also use negative log likelihood cost function, which will yield similar optimization result after our test.

Algorithm \ref{alg:BayesOpt} summarizes the \BO{} procedure for Kalman filter tuning. 
 An attractive feature of \BO{} is that it naturally provides uncertainty quantification on the shape of the objective function at both sampled and unsampled locations. This allows \BO{} to cope with multiple local minima in the parameter space $\bovarspace$. 


{\centering
\begin{minipage}{.92\linewidth}

 \begin{algorithm}[H]
            \caption{\BO{} for Kalman filter tuning}
            \label{alg:BayesOpt}
            \begin{algorithmic}[1] 
                    \State Initialize TP with seed data $\left\{\bovar_s, \fy_s\right\}_{s=1}^{N_{seed}}$ and hyperparameters $\Theta$
                    \While {termination criteria not met}
                        \State $\bovar_j = \argmax_{\bovarspace} \fa$ 
                        \State Evaluate $y(\bovar_j)$, e.g. using $\Jnees(\bovar)$ or $\Jnis(\bovar)$. 
                        \State Add $y(\bovar_j)$ to $\fv(Q)$, $\bovar_j$ to $Q$, and update $\Theta$ \label{ln:testing}
                    \EndWhile \\
                    \Return $\bovar^* = \arg \min_{\bovar_j \in Q} \fv(\bovar_j)$
            \end{algorithmic}
        \end{algorithm}
\end{minipage}
\par
}

\begin{figure*}[htbp!] 
    \centering
	\includegraphics[width=180mm]{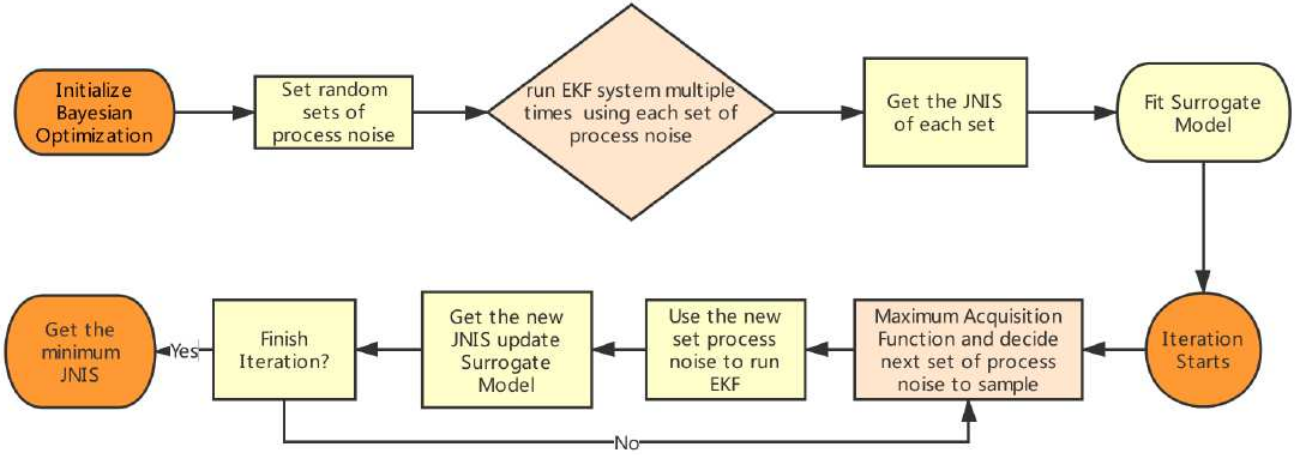}
    \caption{System Flowchart: Bayesian optimization will fit a surrogate model based on some sampled input parameters and associated costs according to the ``black box'' objective function, and then starts iterating until a minimum cost is found.
    }  \label{PG1} 
\label{xfigDiagram}
\end{figure*}

\Ignore{
 As a `black box' optimization method, Bayesian optimization requires a
(possibly stochastic) objective (or cost) function and an input parameter vector
to optimize. In our problem, the input parameter vector will be the process noise
parameters and the objective function will be tied to the EKF NIS or NEES
statistics. Our objective function may be tied to the NIS statistic to
avoid requiring ground truth; the objective function tied to NEES can be viewed
in \cite{chen2018weak}. If the dynamical consistency conditions are met,
then it is easy to show that NIS is $\chi^2$ random variables
\cite{Bar-Shalom2001}. Thus, we need use multiple Monte Carlo runs of the EKF
system to get NIS statistics via simulation or live data collection runs, which
can then be used to define a cost function. To get the cost via simulation, we can simulate some measurements and then add additive white noise to the simulated measurements, finally we use Eq.\  \ref{eq:nisDef}, \ref{eq:nisAvgk} and \ref{JNIS_def} to calculate the cost. 
}
Based on samples over the system input and
corresponding outputs from the objective function,
Bayesian optimization fits a surrogate model of the ``true''
objective function. The optimization method then repeats this process to
find a minimum of the current surrogate model, update this surrogate
model, and find the minimum of the new surrogate model until pre-set
termination conditions are met. The key steps for the EKF-Bayesian optimization
tuning procedure are shown in Figure \ref{PG1}.\\

After the iteration starts, we can see in the flow chart that we maximize
the acquisition function. The new point that can maximize the acquisition function will be chosen as the next set of
process noise to run the EKF system. After running the EKF system with the new
set of process noise we can get a new cost, which will be used to update the
surrogate model.

\section{Application to Extended Kalman Filters}
\label{sct:result}

\indent We evaluated our Bayesian optimization auto-tuning algorithm on a nonlinear state estimation application that uses the Extended Kalman filter (EKF). This application is a closed-loop control system for the aero-robotic Mars Science Lab (MSL) Skycrane landing system. In this case, TPBO is used to automatically tune the assumed process noise covariance matrix, which is one of the main ``tuning knobs'' for the EKF.
The ``Skycrane maneuver'' was used as a deployment method for the Mars Science Laboratory (MSL) Curiosity rover upon its arrival and descent near the surface of Mars, as an alternative to the air bag method used in previous missions. Thrusters were used to stabilize the MSL Descent Stage System (a robotic aircraft) to zero horizontal velocity and to slowly guide the system to 20m above the ground to deploy the rover. A simplified model of this latter stage’s longitudinal dynamics will be used to simulate vehicle state estimation just prior to the rover deployment phase. Figure \ref{fig:skycrane} depicts a simplified 2D longitudinal dynamics model of the MSL Descent Stage aircraft. More detailed descriptions of the MSL platform are given in \cite{steltzner2006mars}, \cite{mitcheltree2006mars}.\\

\subsection{System description}
The system is modeled as a rectangular box with two thrusters, one each on the bottom corners of the aircraft mounted at angle $\beta$ to the vehicle $z$ axis. The simplified vehicle states consist of the inertial translation $\xi$ (m), altitude above surface $z$ (m), pitch angle $\theta$ (radian), and rates $\dot{\xi}$ (m/s), $\dot{x}$ (m/s), and $\dot{\theta}$ (rad/s). The control inputs 
are defined in terms of the thrusts $T_i$ (Newtons) produced by the $i^{th}$ thruster. The state and control input are therefore 
\begin{equation} \label{sim_motion_model}
    \begin{split}
    \mathbf{x}(t) &= (\xi,\dot{\xi}, z, \dot{z},\theta,   \dot{\theta})^T \\
    \mathbf{u}(t) &= (T_1, T_2)^T.
    \end{split}
\end{equation}

\begin{figure}
    \centering
    \includegraphics[width=85mm]{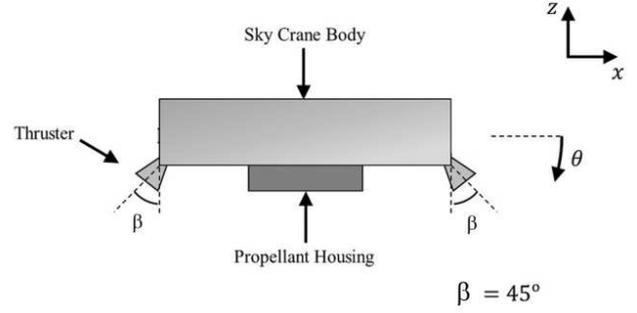}
    \caption{The Skycrane aircraft has two thrusters angled at 45 degrees, which nominally try to keep the platform 20 meters above the ground with zero translational and rotational motion.}
    \label{fig:skycrane}
\end{figure}

The equations of motion are derived here by considering only gravity, thrust, and drag forces (the vehicle is assumed not to generate significant lift in this phase). Drag will be modeled as $F_{drag} = 1/2C_D\rho A v^2$, where $C_D$ is the drag coefficient, $\rho$ is the atmosphere density, $v$ is the magnitude of the velocity, and $A$ is the approximate cross-sectional area of the vehicle in the direction of motion. Let $m_b$ be the mass of the Skycrane aircraft and payload, $m_f$ be the mass of the fuel, $\omega_b$ and $h_b$ the width and height dimensions of the Skycrane body as shown in Fig. \ref{fig:skycrane}, $\omega_f$ and $h_f$ the dimensions of the propellant housing, and $h_{cm}$ and $\omega_{cm}$ the dimensions for the vehicle center of mass. The motion equation are written as\\
\begin{equation} \label{skycrane_pr_model}
    \begin{split}
    \ddot{\xi} &= \frac{T_1(\sin(\theta + \beta)) + T_2(\sin(\theta - \beta)) - F_{D,\xi}}{m_b+m_f} + \widetilde{\omega}_1 \\
    \ddot{z}  &= \frac{T_1(\cos(\theta + \beta)) + T_2(\cos(\theta - \beta)) - F_{D,\xi}}{m_b+m_f} - g + \widetilde{\omega}_2 \\
    \ddot{\theta} &= \frac{1}{I_\eta}((T_1 -T_2)(\cos\beta \frac{\omega_b}{2} - \sin\beta h_{cm}) ) +  \widetilde{\omega}_3 \\
    \frac{1}{I_\eta} &= \frac{1}{12}(m_b(\omega_b^2 + h_b^2)+ m_f(\omega_f^2 + h_f^2))\\
    F_{D,\xi} &= \frac{1}{2}C_D \rho (A_{s}\cos(\theta - \alpha) + A_{b}\sin(\theta - \alpha)) \dot{\xi} \sqrt{\dot{\xi^2} + \dot{z^2}} \\
    F_{D,z} &= \frac{1}{2}C_D \rho (A_{s}\cos(\theta - \alpha) + A_{b}\sin(\theta - \alpha)) \dot{z} \sqrt{\dot{\xi^2} + \dot{z^2}}\\
    \alpha &= \tan^{-1}\frac{\dot{z}}{\dot{\xi}}
    \end{split}
\end{equation}
To simplify the model further, changes in $m_f$ will be ignored. Values for these constants are provided in the appendix.

\indent Sensors for state estimation consist of a simplified ideal single-axis IMU, i.e. an accelerometer and rate gyro pair which provide noisy measurements of inertial $\xi$ accelerations and pitch rotations about the inertial $\zeta$ axis.  

\begin{figure}
    \centering
    \includegraphics[width=95mm]{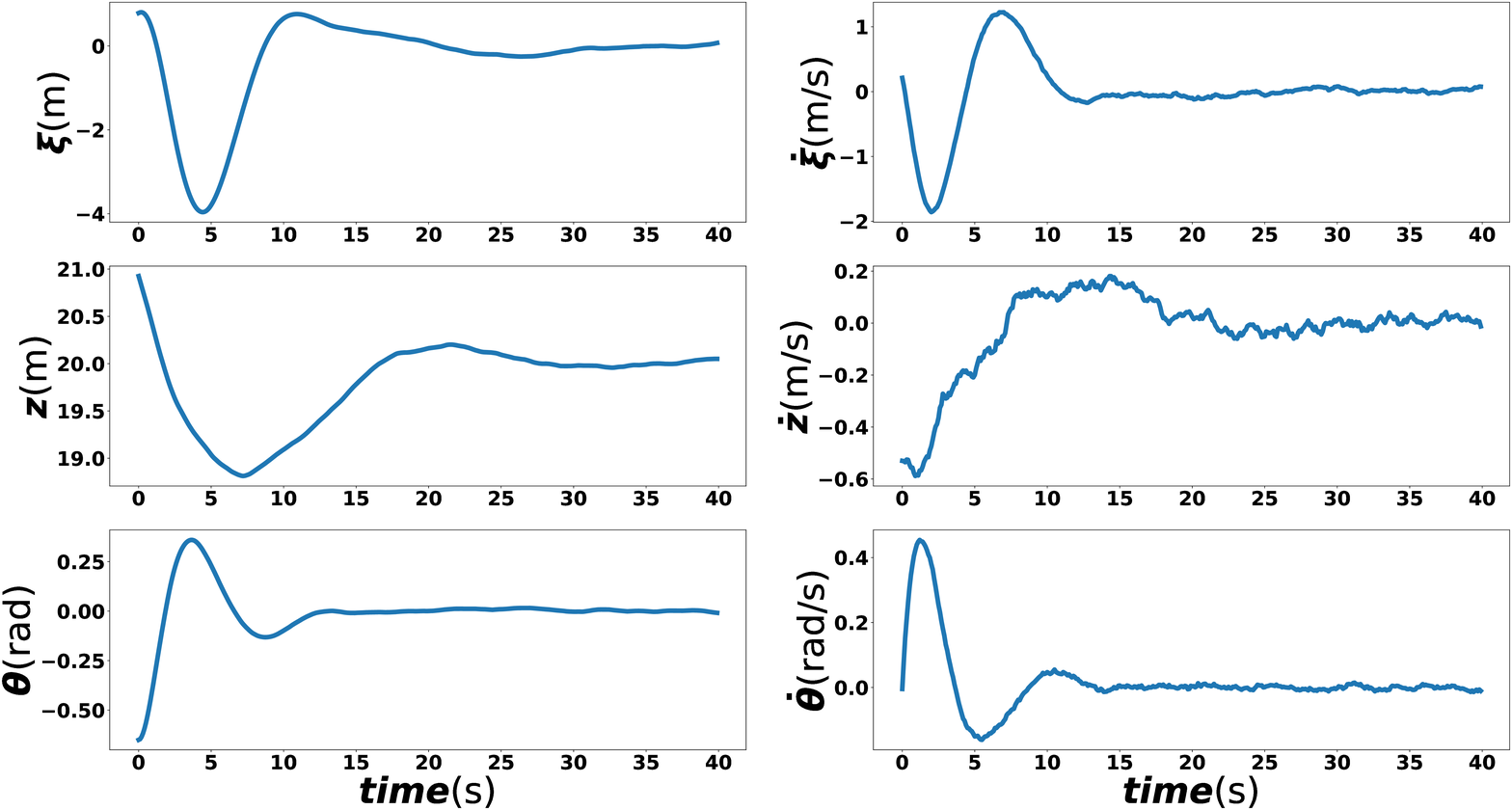}
    \caption{Sample Skycrane simulation showing the LQR controller maintains the desired reference state using EKF-estimated full-state feedback. 
    }
    \label{fig:skycrane_state_his}
\end{figure}

The sensor data also include on-board
barometer readings to gauge altitude. Image-based tracking measurements from an overhead passing satellite are also converted into noisy $\xi$ platform position reports. The measurement vector can be written as \\
\begin{equation} \label{skycrane_me_model}
    \mathbf{y} = 
        \begin{bmatrix}
            \xi \\
            z   \\
            \dot{\theta} \\
            \ddot{\xi}
        \end{bmatrix}
        +
        \widetilde{\mathbf{v}}(t)
\end{equation}
where $\widetilde{\mathbf{v}}(t) \in \mathbb{R}^4$ is the sensor error vector. The process disturbance and measurement noise vectors are
\begin{align} 
    \widetilde{\mathbf{w}}(t) &= (\widetilde{\omega}_1, \widetilde{\omega}_2, \widetilde{\omega}_3)^T, \label{noise_skycrane_w}\\
    \widetilde{\mathbf{v}}(t) &= (\widetilde{v}_1, \widetilde{v}_2, \widetilde{v}_3, \widetilde{v}_4)^T, \label{noise_skycrane_v}
\end{align}
all of which are modeled as additive white Gaussian noise. 
To obtain the appropriate matrices for the EKF, the discrete time state transition matrix is approximated by taking the Jacobian of the Euler-intergrated continuous time motion model $\dot{\mathbf{x}}(t) = f(\mathbf{x}(t), \mathbf{u}(t))$ (with sample period $\delta t=0.1$s). The Jacobian of measurement model is obtained from Eq.\ (\ref{skycrane_me_model}). One important thing we need to notice is that our system now is in continuous time, to implement it we need convert it into discrete time, which needs some extra work. The details for obtaining the corresponding Jacobian matrices and discretization are provided in the Appendix. 
%
Note that the vehicle must maintain a desired nominal trim state of $\mathbf{x}_{ref} = (0,0,20,0,0,0)^T$ (steady hover 20 m above the surface). Linearization about the trim state reveals that the continuous time perturbation dynamics are unstable but controllable and observable.  
Hence, to maintain the platform at the desired state using estimated full-state state feedback, a Linear Quadratic Regulator (LQR) controller is also used to define the control inputs $\mathbf{u}(t)$ at each discrete time step according to the control law,  
\begin{equation}
    \mathbf{u}_k = \mathbf{u}_{nom} - \mathbf{K}_{lin}(\mathbf{x}_{k} - \mathbf{x}_{ref})
\end{equation}
where the $\mathbf{K}_{lin}$ is a pre-calculated LQR gain, which can be obtained offline using the separation principle assuming ideal full-state feedback for the linearized dynamics about trim (values given in Appendix). The same closed-loop control law is used throughout the Bayesian optimization auto-tuning procedure and the thrust values are made available to the EKF. 
The nominal thrusts $\mathbf{u}_{nom}$ correspond to when the aircraft stabilizes to the desired state without process noise, and is given by
\begin{equation}
    T_{1,nom} = T_{2,nom} = 0.5g\frac{m_b+m_f}{\cos\beta}.
\end{equation}
\indent An example of running the EKF for the Skycrane system can be seen in Figure \ref{fig:skycrane_state_his}, which shows the EKF's estimated state values over time with the help of LQR controller. Each element's variance of the process noise is $(0.01, 0.01, 0.001)$ for $\widetilde{\mathbf{w}}(t)$ and variance of measurement noise is $ (1.0, 0.5, 0.025, 0.0225)$ for $\widetilde{\mathbf{v}}(t)$. They all have zero mean. 

\subsection{Discrete EKF From Continuous Time Model}
The EKF prediction stage's formula from the continuous time will be different from the general discrete time EKF. \\
The prediction stage is 
\begin{align}
    \xCond{k}{k-1} &= f(\xCond{k-1}{k-1}, \uVec{k}), \label{prediction2_1}\\
    \covCond{k}{k-1} &=  {\widetilde{\mathbf{F}}_k} \covCond{k-1}{k-1}{ {\widetilde{\mathbf{F}}_k^\top}} + \mathbf{\Omega}_k\Q{k}\mathbf{\Omega}_k^T, \label{prediction2_2}
\end{align}
We cannot directly obtain $f(\xCond{k-1}{k-1}, \uVec{k})$ using the continuous time formula. There are some other ways. The first method is that we can use the first order linearized form of $f(\xCond{k-1}{k-1}, \uVec{k})$ to estimate it, which means
\begin{equation}
    f(\xCond{k-1}{k-1}, \uVec{k}) \approx \xCond{k-1}{k-1} +  \delta t f(\xCond{k-1}{k-1}, \uVec{k})
\end{equation}
The second method is that we can numerically solve the ordinary differential equations (ODE) $f(\xCond{k-1}{k-1}, \uVec{k})$, which can yeild a more precise result. In our implementation we use the ODE integration library \cite{ahnert2011odeint} to estimate the Eq.\  \eqref{prediction2_1}. In equation \eqref{prediction2_2} $\widetilde{\mathbf{F}}_k \approx \mathbf{I} + \delta t \mathbf{F}_k$, $\mathbf{F}_k$ can be computed from Eq.\  \eqref{process_jacobian}. $\mathbf{\Omega}_k \approx \delta t \mathbf{\Gamma}_k$. $\mathbf{\Gamma}_k$ is a mapping matrix from the 3 dimension noise to the 6 dimension noise, which can be seen in the Appendix. The process noise noise covariance is \\
\begin{equation}
    \Q{k} = 
    \begin{bmatrix}
        \Q{\ddot{\xi}} & 0 & 0 \\
        0 & \Q{\ddot{z}}   & 0 \\
        0 & 0 & \Q{\ddot{\theta}}         
    \end{bmatrix}
\end{equation}
The update stage will remain the same as Eq.\ \ref{update1_dis1} to \ref{update1_dis4}. The measurement noise and its covariance are still fixed and written in the Appendix.

\subsection{Optimization results}
 As a first simple experiment, \BO~ is used to perform a 1D parameter search for $\Q{k}$ defined as a constrained diagonal matrix, where the process noise covariances are such that $\Q{\ddot{\xi}} = \Q{\ddot{z}} = 10*\Q{\ddot{\theta}}$. 

\begin{figure*}
    \centering
  \subfloat[iteration1]{%
       \includegraphics[width=0.31\textwidth]{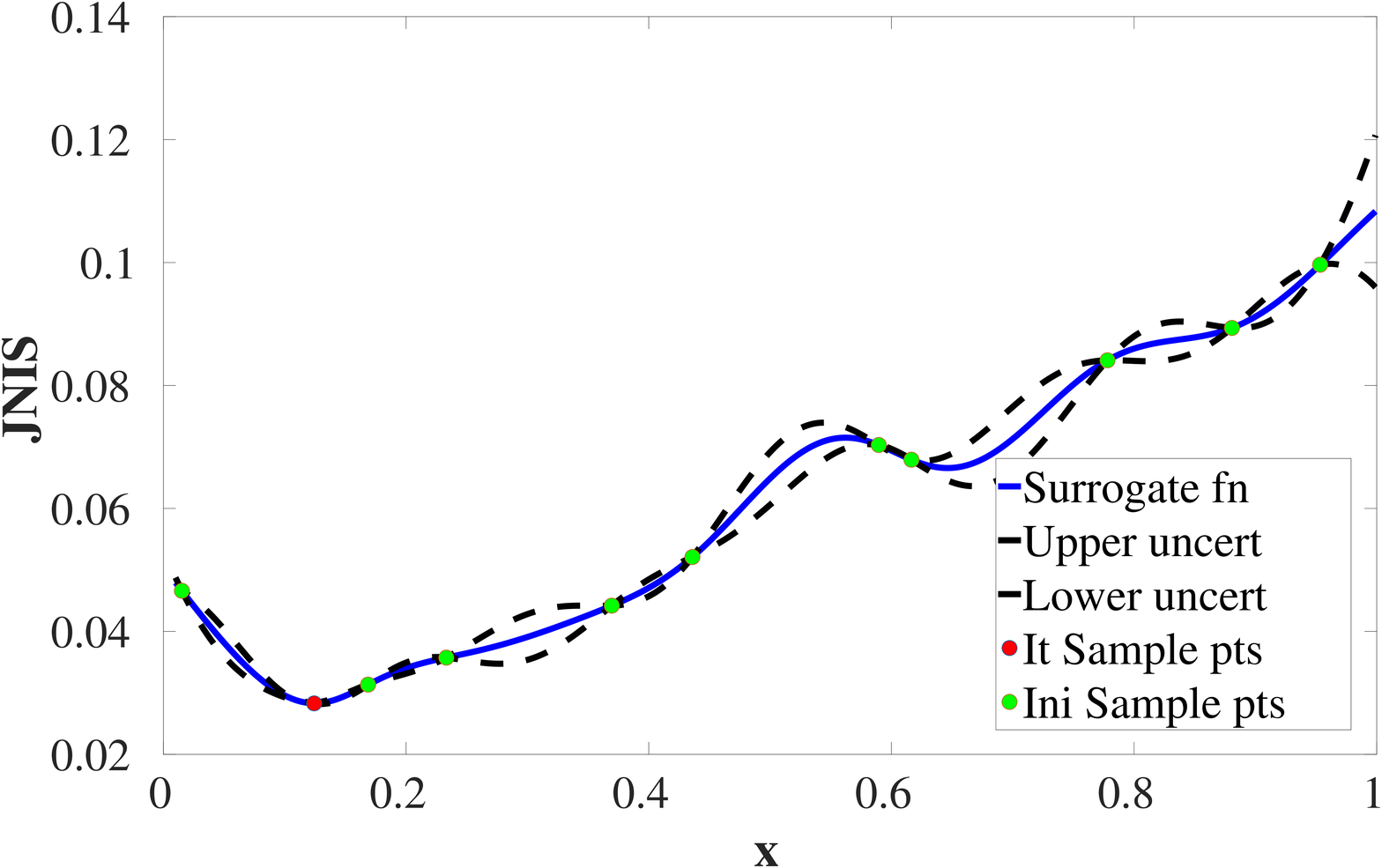}}
    \label{sky1dit1} 
  \subfloat[iteration10]{%
        \includegraphics[width=0.31\textwidth]{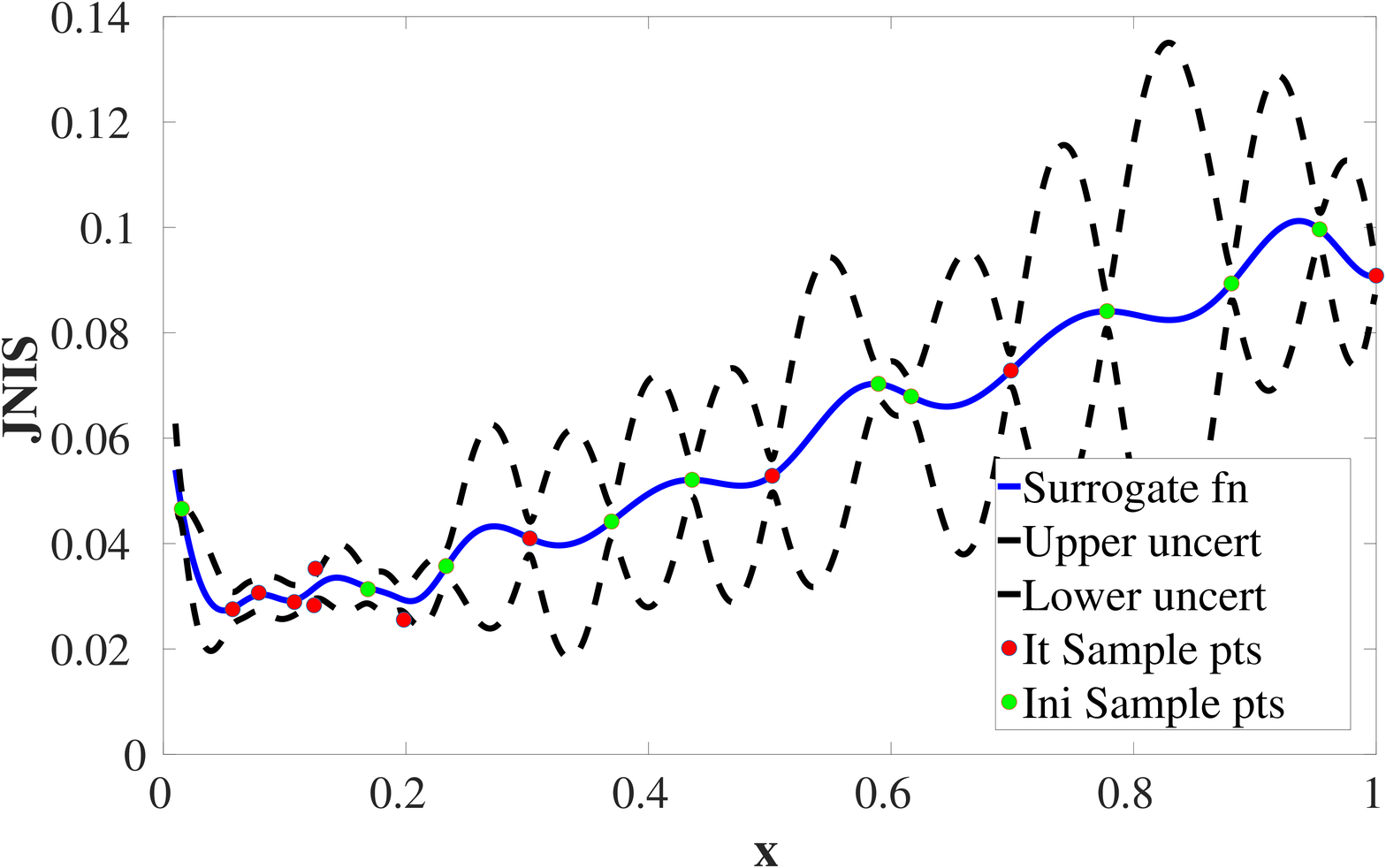}}
    \label{sky1dit10}
  \subfloat[iteration50]{%
        \includegraphics[width=0.31\textwidth]{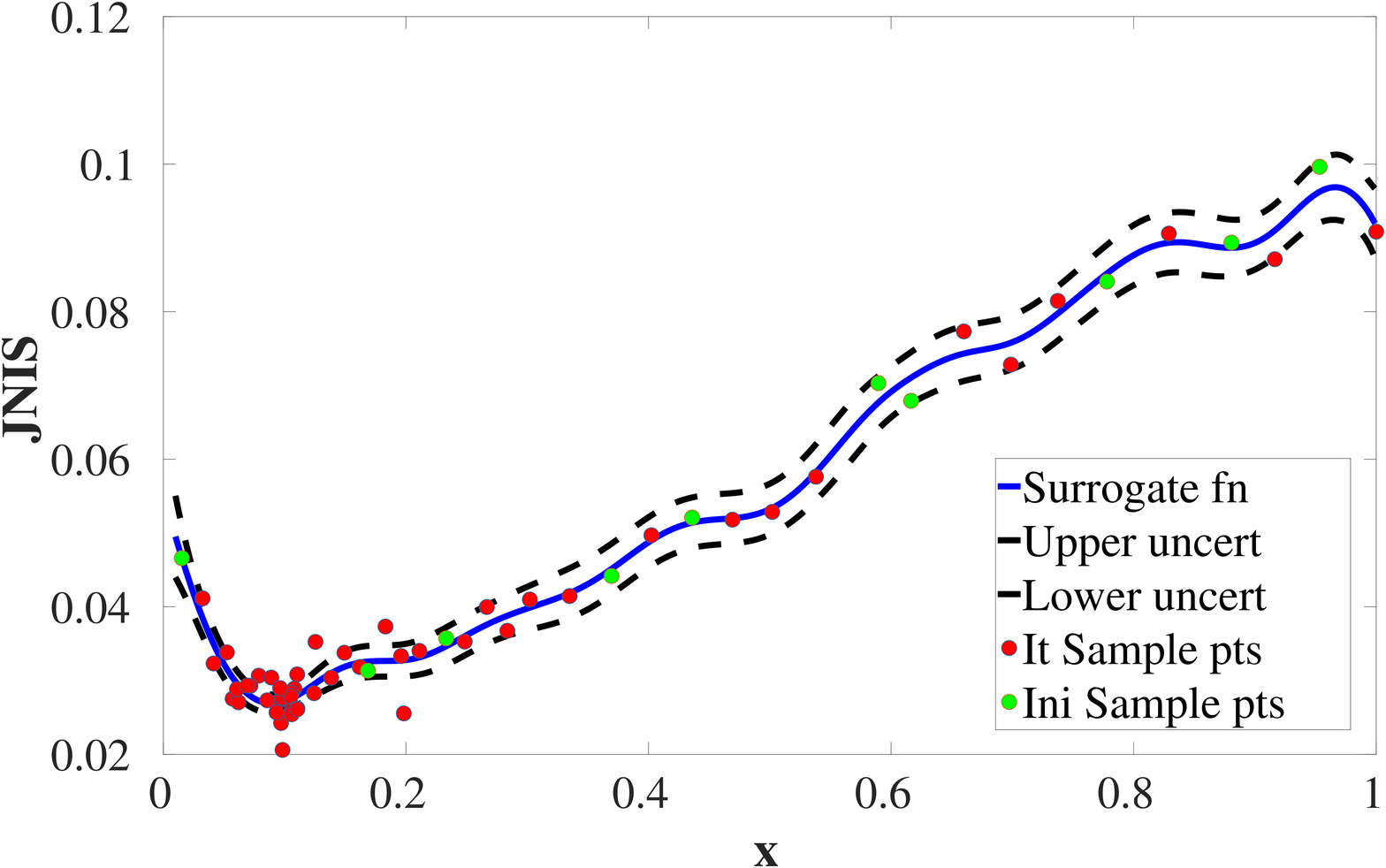}}
    \label{sky1dit50}
  \caption{1D plots of surrogate functions and uncertainty bounds for the Skycrane problem, as well as sample points chosen by Bayesian optimization. The black dashed line represents the lower and upper uncertainty bounds for a 95\% confidence interval. The green dots are the initial sample points. The red dots are the sample points after the iteration starts. The blue line is the mean of the surrogate model} \label{fig:surrogate_model_1d_skycrane}
\end{figure*}

For the 1D parameter optimization, TPBO was applied over the range $\Q{\ddot{\xi}} \in [1 \times 10^{-2}, 1]$, using 10 initial surrogate model seed samples, 50 total iterations and 200 Monte Carlo run. The surrogate model and samples points for different sample iterations are shown in Figure \ref{fig:surrogate_model_1d_skycrane}. Table \ref{table:result_skycrane} also shows the numerical values for the final best minimizer found. 

\begin{figure*}
    \centering
  \subfloat[iteration1]{%
       \includegraphics[width=0.31\textwidth]{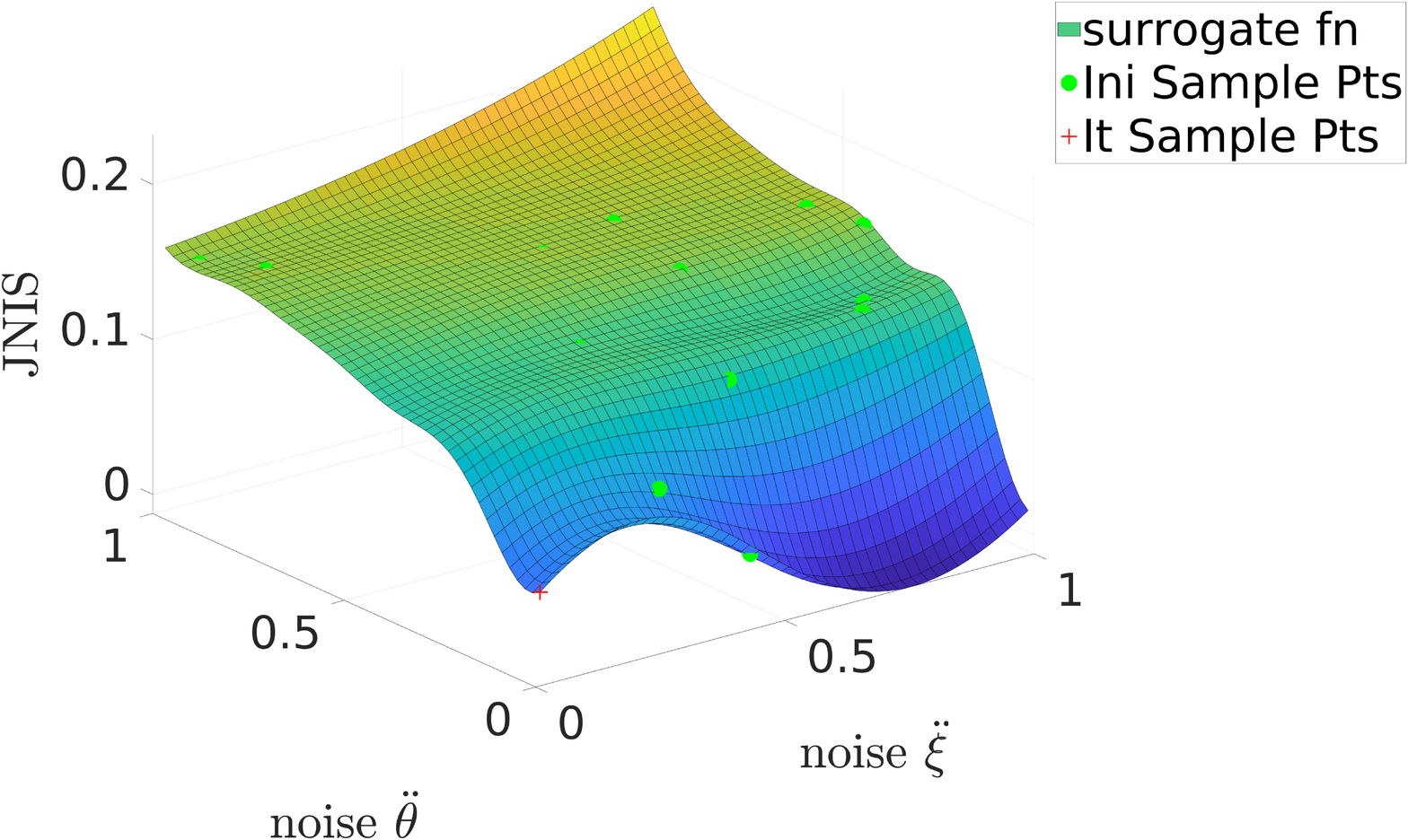}}
    \label{sky2dit1} 
  \subfloat[iteration50]{%
        \includegraphics[width=0.31\textwidth]{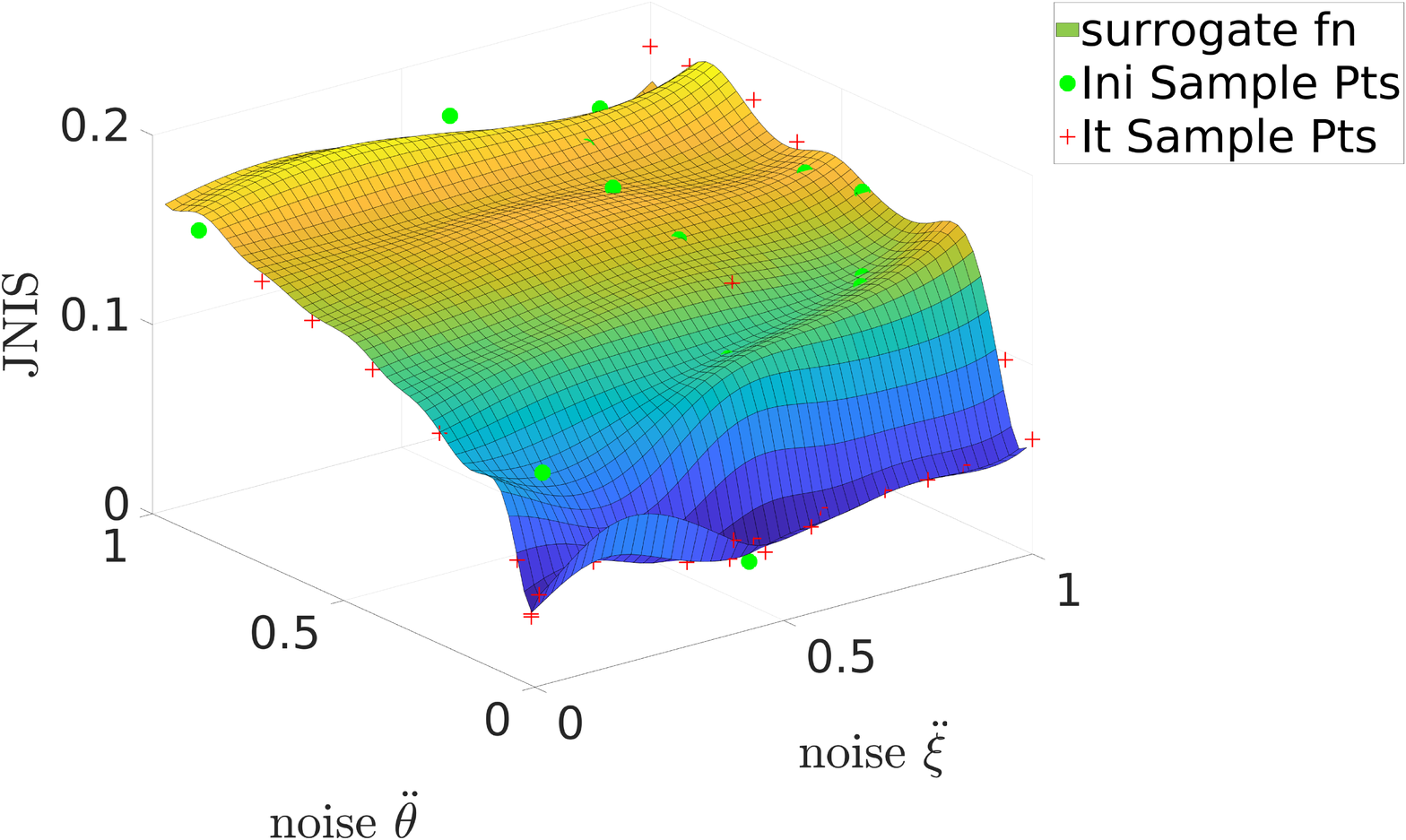}}
    \label{sky2dit50}
  \subfloat[iteration80]{%
        \includegraphics[width=0.31\textwidth]{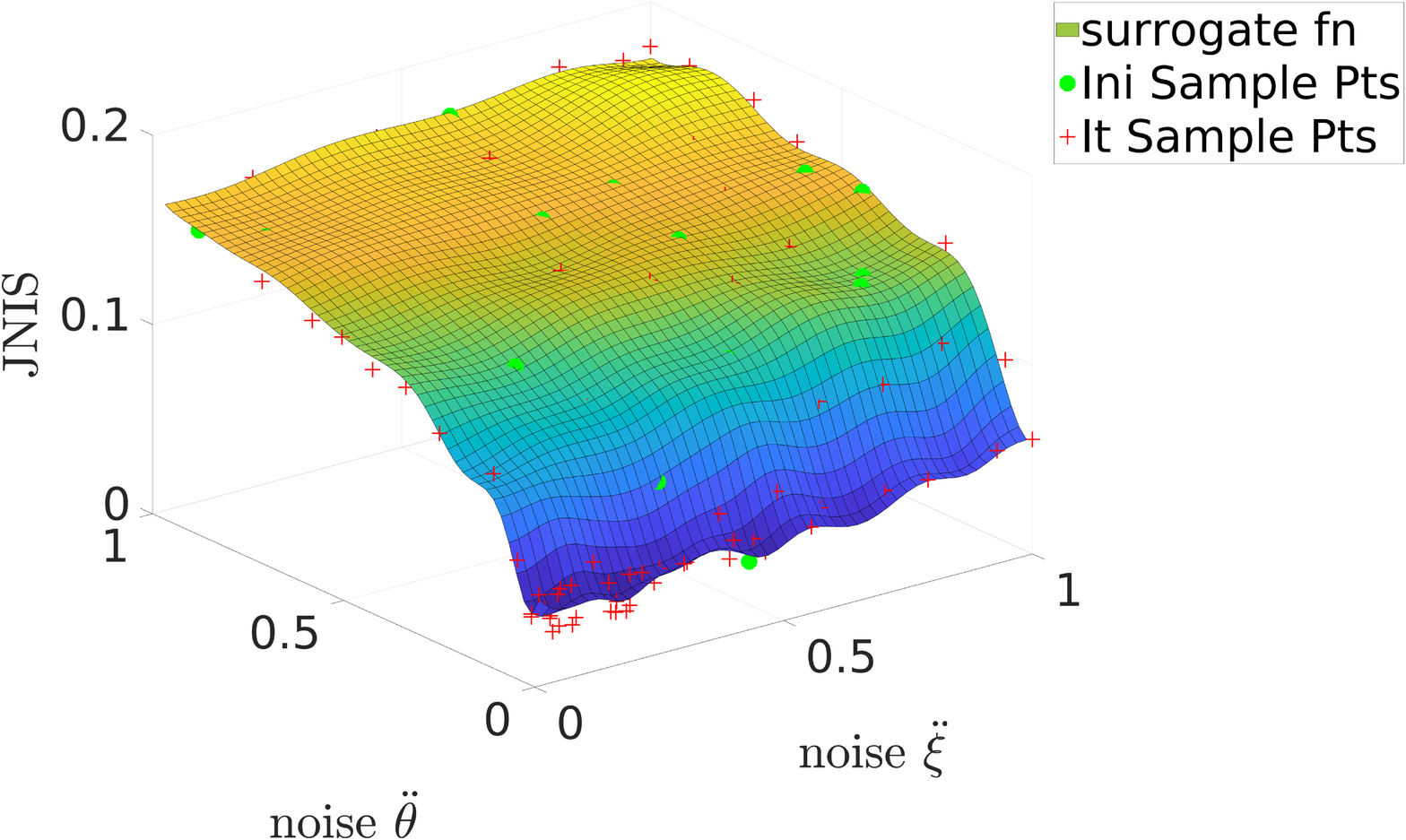}}
    \label{sky2dit80}
  \caption{2D surface plots of surrogate functions for the Skycrane problem, as well as sample points. The green circles are the initial sample points and the red cross is the sample points after the iterations starts. Upper and lower bound surfaces are not plotted here or the surfaces may cover each other.} \label{fig:surrogate_model_2d_skycrane}
\end{figure*}

For the 2D parameter optimization, the parameter $\Q{\ddot{z}} = 0.1$ is held fixed, while $\Q{\ddot{\xi}}$ and $\Q{\ddot{\theta}}$ are optimized. 
The lower bound and upper bound are set as $[1 \times 10^{-2}, 1 \times 10^{-3}]$ to [1,1] respectively. We have 20 initial samples, 80 iterations and 200 Monte Carlo run. Again, the mean value of the surrogate model and the sample points at different iterations the \BO{} found are shown in Figure \ref{fig:surrogate_model_2d_skycrane}. From 1D optimization result we can clearly see the result converge to points around 0.1 with high confidence and from the 2D optimization result we can see the result converge to points around [0.1, 0.01]. However, from the 2D result Table  \ref{table:result_skycrane} we can see after 80 iterations, the cost does not change significantly as the $\Q{\ddot{\xi}}$ change when the $\Q{\ddot{\theta}}$ is around 0.1. This phenomenon may lead to a non-optimal result from TPBO. In Bayesian optimization, this happens when certain dimensions do not have a great impact on the cost, which encourages the addition of weights on other dimensions.

\begin{figure*}
    \centering
  \subfloat[RMSE of $\xi$]{%
       \includegraphics[width=0.31\textwidth]{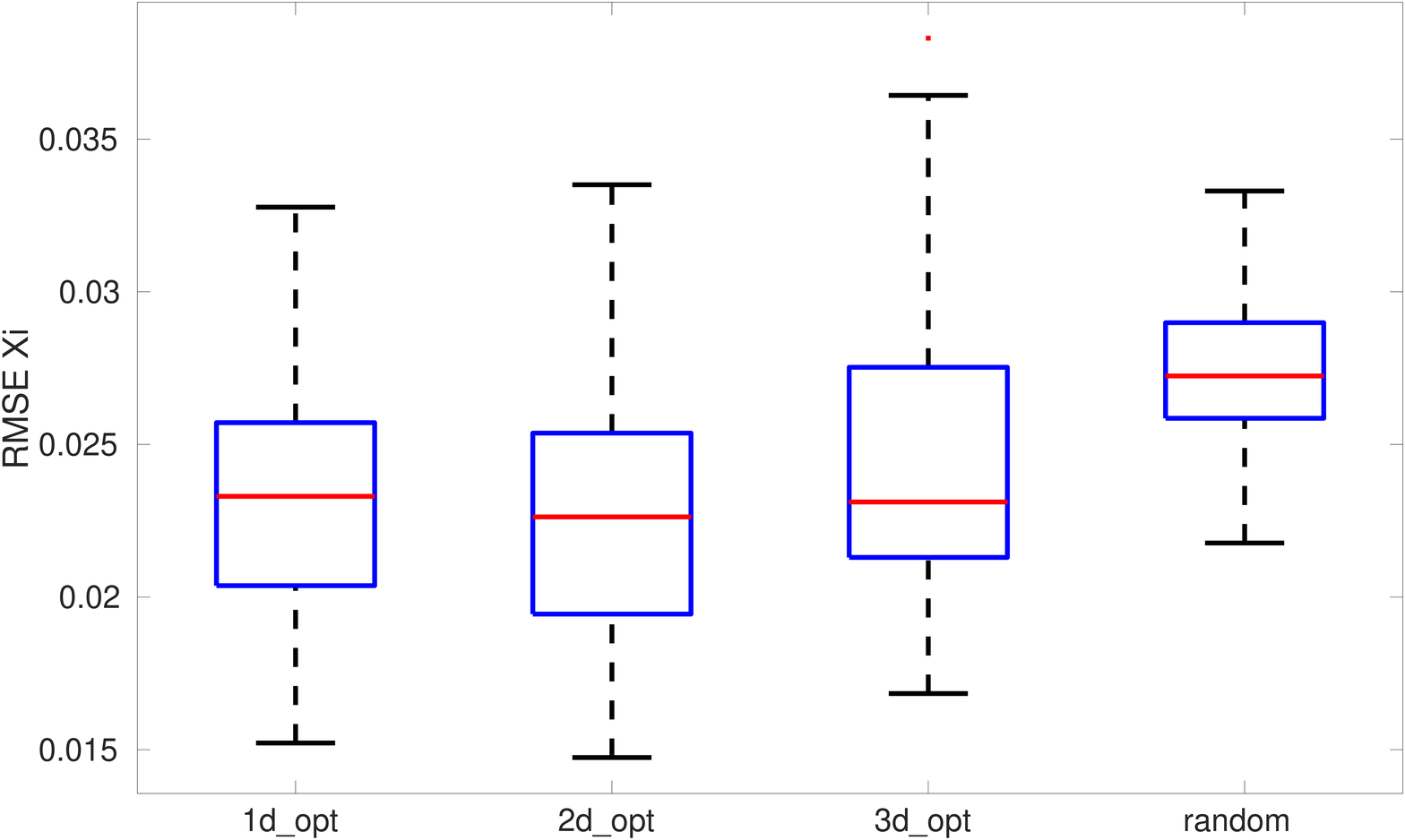}}
    \label{rmse_xi} 
  \subfloat[RMSE of $z$]{%
        \includegraphics[width=0.31\textwidth]{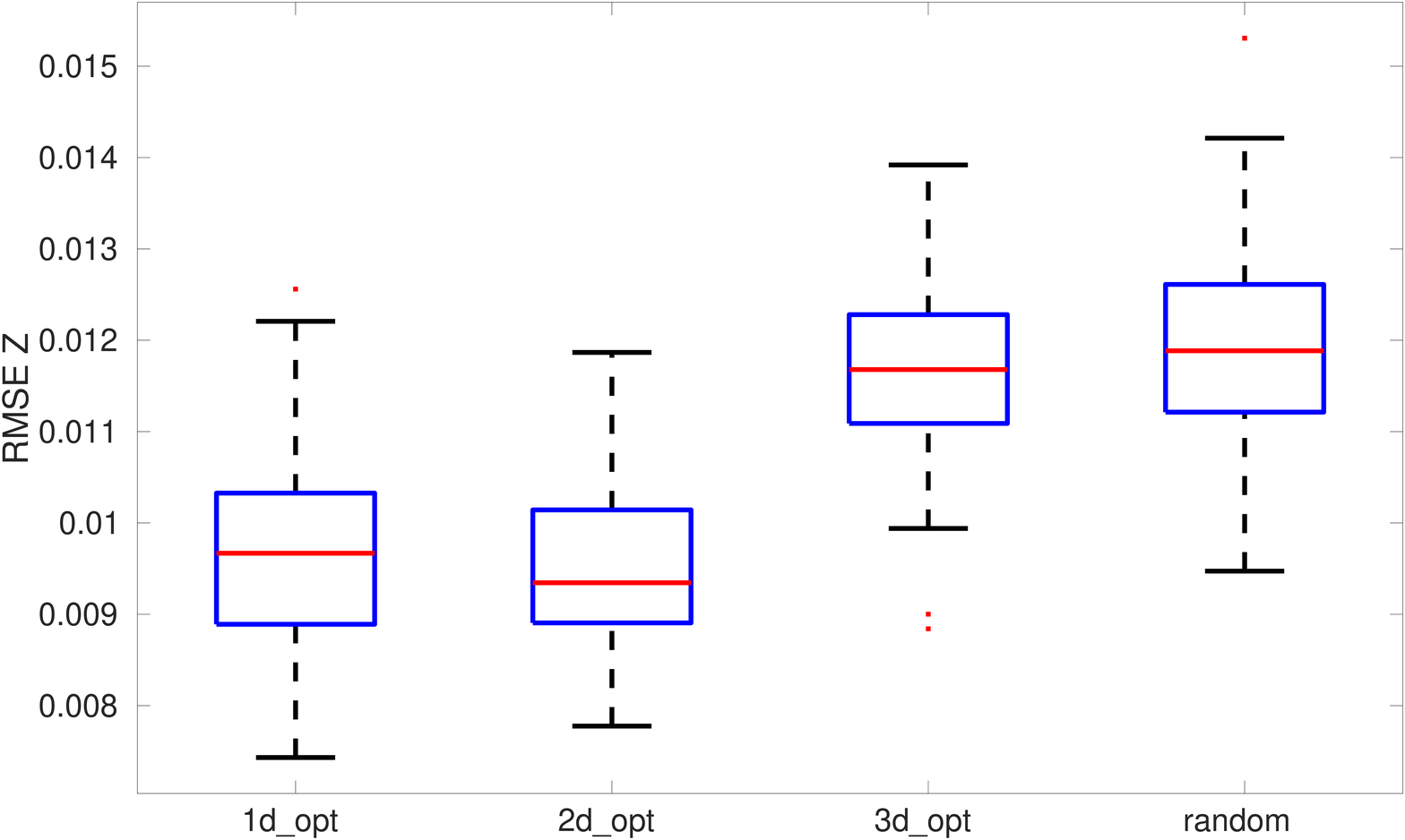}}
    \label{rmse_z}
  \subfloat[RMSE of $\theta$]{%
        \includegraphics[width=0.31\textwidth]{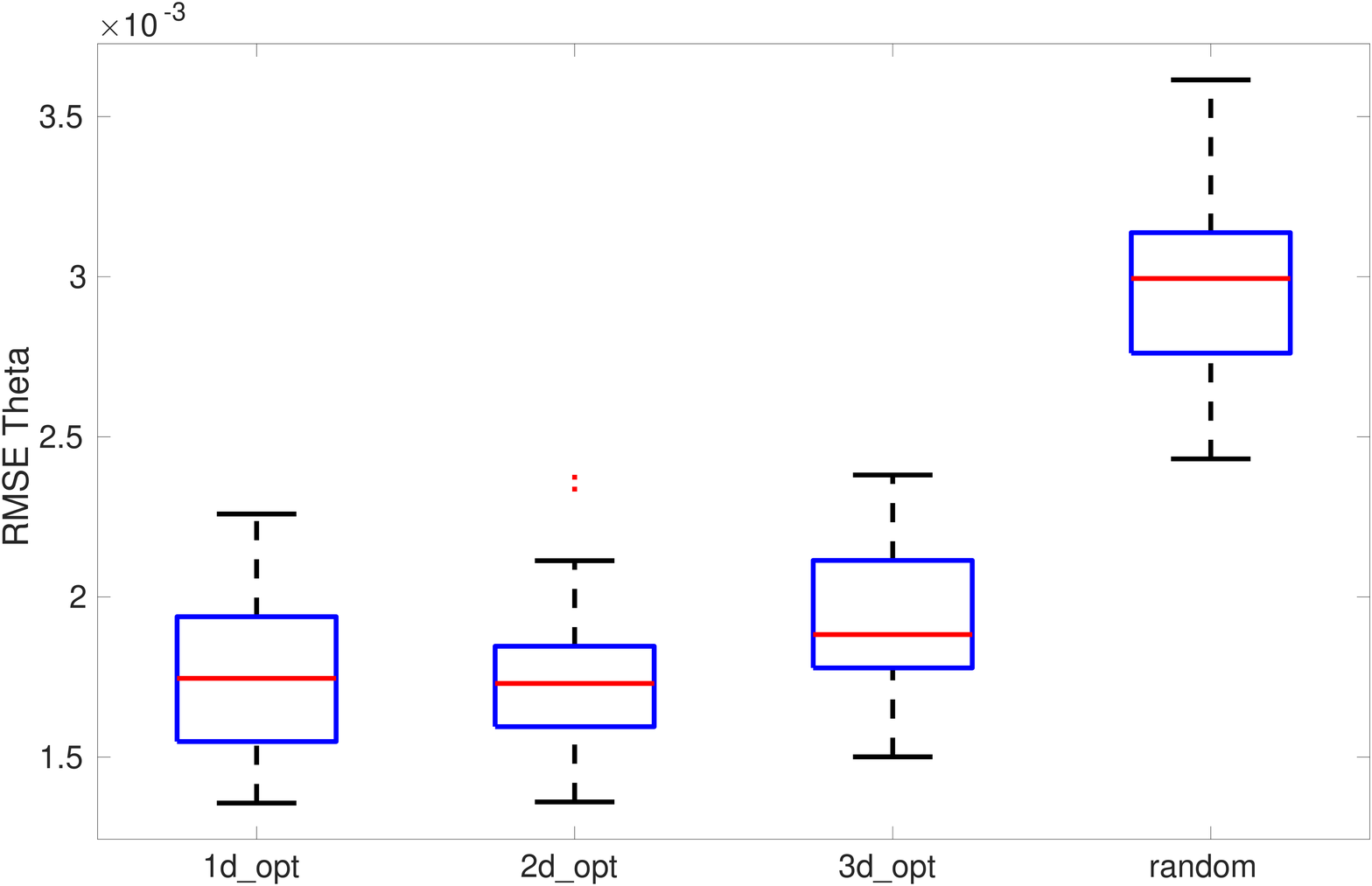}}
    \label{rmse_theta}
  \caption{\label{fig:rmse_skycrane} \texttt{Nd\_opt} means we use the optimization result from the $N$D search and run Skycrane system. The left one shows the RMSE of translation and the right one shows the RMSE of rotation.} 
\end{figure*}

For the 3D optimization result, the boundaries for $\Q{\ddot{\xi}}$, $\Q{\ddot{z}} $, $\Q{\ddot{\theta}}$ are $[1 \times 10^{-2}, 1 \times 10^{-2}, 1 \times 10^{-3}]$ to $[1,1,1]$ respectively. We have 30 initial samples, 100 iterations and 200 Monte Carlo run for each sample. The value of $\Q{\ddot{z}}$ is far away from the optimal and it suffers from the same reason as the 2D optimization, which yield a relative larger error when we check the RMSE of $z$ in Figure \ref{fig:rmse_skycrane}. 
\begin{figure}[H]
    \centering
    \includegraphics[width=95mm]{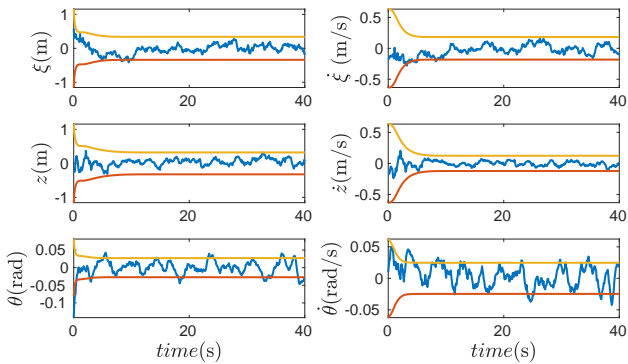}
    \caption{\label{fig:consistency_check_skycrane} An example time series error between Skycrane estimates using the BO-tuned optimal parameter values and ground truth states (orange lines: $2\sigma$ bounds; blue line: error). Intuitively, the error should lie between the $2\sigma$ bounds about 95\% of the time if the filter is consistent.}
\end{figure}


To validate the results of TPBO, the RMS error of three states $\xi, z,\theta$ between the EKF's state estimation and the groundtruth (state from the simulator) are evaluated. 
We apply the 1D, 2D, 3D optimized parameters with a random set of process noise for 50 simulation runs of the Skycrane EKF and then obtain results in Figure \ref{fig:rmse_skycrane}. These results show that the 2D optimization results yield the best state estimates, since it has the minimum median error and smallest lower and upper error bounds. 
The 1D optimization result is similar to 2D optimization result. 
It is worth noting that all the errors are in fact small; for example, even though the RMSE of $\theta$ is visually larger than the others, its median value is $3\times 10^{-3}$ rads. This is because 
measurement noise is set to fit the model precisely, so that its behavior will be robust most of the time. Under this condition, \BO{}'s optimization ability is assessed over a relatively small range of NIS cost values. Figure \ref{fig:consistency_check_skycrane} shows a typical trace for the state estimation error using the 3D optimized parameter estimates, indicating that consistent estimates are in fact obtained.\\
\Ignore{
this\\
can\\
delete\\
the \\
space\\
}
\begin{table}
  \begin{center}
  \begin{tabular}{ |c|c|c|}
    \hline 
    \diagbox{Type}{Result}  & Cost & Optimal  \\
    \hline
    {1D opt} &0.0206 &(0.098,0.098,0.0098)  \\      
    \hline 
    {2D opt} &0.0251 &(0.0446,0.1,0.0119)  \\
    \hline
    {3D opt} &0.0193 &(0.0349838, 0.999984,0.0152815)  \\
    \hline
  \end{tabular} 
  \end{center}
  \caption{\label{table:result_skycrane} \upshape Optimal means the optimal process noise for the EKF covariance. They stands for $\Q{\ddot{\xi}}$, $\Q{\ddot{z}} $, $\Q{\ddot{\theta}}$ respectively. } 
\end{table}

\section{Conclusion}
\label{sct:final}
As a black box optimization method, \BO{} simplifies what we need to know about a system in order to get the minimum cost. We used an example, Skycrane State estimation to show that this algorithm can be applied to complex nonlinear systems. This novel approach can also help the practitioners get the optimal process noise covariance much faster than tuning the EKF manually. In this paper we have shown results using an NIS-based cost function only. Although a NEES-based cost function can also work just as well, as shown in ref. \cite{chen2018weak}, this requires the availability of ground truth state information, so NIS-based cost functions may often be more practical and are applicable with real sensor data. In this paper, we also have focused only on optimizing filter process noise parameters. However, the same auto-tuning process can be applied if the measurement noise also needs to be adjusted for a particular application \cite{chen2018weak}. In fact, if we don't have confidence in either process noise or measurement noise covariances, it is possible to use \BO{} to optimize these parameters simultaneously. The flexibility of the \BO{} allows us to do more.\\


\indent In the future, as this algorithm is robust to use, we aim to apply it to more realistic hardware-based system tuning problems. Another interesting direction for future research involves optimization of higher dimensional parameter spaces, where some dimensions may potentially have little/no noticeable effect on the NIS tuning cost. Possible strategies for handling this might include using TPBO to optimize those parameters which have a significant impact on the tuning cost, leaving the remaining parameters to be hand-tuned. Since the TPBO algorithm is able to support arbitrary ``black box" cost function evaluations, modifications or alternatives to the NIS cost function could also be explored. 
Finally, the flexibility of our \BO{} approach means that it can be applied to other optimal estimator tuning problems. Most notably, for example, we have already applied this approach to VI SLAM (Visual Inertial Simultaneous Localization and Mapping) extrinsic parameter calibration \cite{chen2018visual}, where the ``extrinsic parameter'' means the relative pose between the camera and other sensors. \\


%

\appendices
\section{Additional Information of Sky-Crane Model}
\subsection{Jacobian and Parameters}
\indent The process model for the Skycrane has the form $\dot{\mathbf{x}}(t) = f(\mathbf{x}(t), \mathbf{u}(t)) + \widetilde{\mathbf{w}}$, where
\begin{equation}
    \begin{bmatrix}
        \dot{\xi} \\
        \ddot{\xi} \\
        \dot{z} \\
        \ddot{z} \\
        \dot{\theta} \\
        \ddot{\theta}
    \end{bmatrix}
    = f\left(
    \begin{bmatrix}
        \xi \\
        \dot{\xi} \\
        z \\
        \dot{z} \\
        \theta \\
        \dot{\theta}
    \end{bmatrix}
    ,
    \begin{bmatrix}
        T_1\\
        T_2
    \end{bmatrix}
    \right)
    +
    \begin{bmatrix}
        0 \\
        \widetilde{\omega}_1 \\
        0 \\
        \widetilde{\omega}_2 \\
        0 \\
        \widetilde{\omega}_3 \\
        0
    \end{bmatrix}
\end{equation}
Substituting for Eq.\ \eqref{skycrane_pr_model} and taking derivatives, the Jacobian of the process model is 
\begin{equation}
    \mathbf{F}(t) = 
        \begin{bmatrix}
            0 & 1 & 0 & 0 & 0 & 0 \\
            0 & \mathbf{F}_{11} & 0 & \mathbf{F}_{13} & \mathbf{F}_{14} & 0 \\
            0 & 0 & 0 & 1 & 0 & 0 \\
            0 & \mathbf{F}_{31} & 0 & \mathbf{F}_{33} & \mathbf{F}_{34} & 0 \\
            0 & 0 & 0 & 0 & 0 & 1 \\
            0 & 0 & 0 & 0 & 0 & 0 
        \end{bmatrix}
\end{equation}
where
\begin{equation}\label{J_detail}
    \begin{split}
        \mathbf{F}_{11} &= nc((A_{sc} + A_{bs})(2\dot{\xi}^2 + \dot{z}^2)/V_t + \dot{\xi}\dot{z}(A_{bc} - A_{ss})/V_t ) \\
        \mathbf{F}_{13} &= nc((A_{sc} + A_{bs})(\dot{\xi}+ \dot{z})/V_t + \dot{\xi}^2(A_{ss} - A_{bc})/V_t) \\
        \mathbf{F}_{14} &= l(T_1 \cos{(\beta + \theta)}  + T_2\cos{(\beta-\theta)} + cd \dot{\xi} V_t(A_{ss} - A_{bc}))\\
        \mathbf{F}_{31} &= nc((A_{sc} + A_{bs})(\dot{\xi}+ \dot{z})/V_t) +  \dot{z}^2(A_{bc} - A_{ss})/V_t)\\
        \mathbf{F}_{33} &= nc((A_{sc} + A_{bs})(2\dot{\xi}^2 + \dot{z}^2)/V_t + \dot{\xi}\dot{z}(A_{ss} - A_{bc})/V_t )\\
        \mathbf{F}_{34} &= l(T_2\sin{(\beta-\theta)}-T_1 \sin{(\beta + \theta)} + cd \dot{\xi} V_t(A_{ss} - A_{bc}))
    \end{split}
\end{equation}
Taking derivates of \eqref{skycrane_me_model},
\begin{equation} 
    \begin{split}
        \mathbf{H}(t) = 
            \begin{bmatrix}
                1 & 0 & 0 & 0 & 0 & 0 \\
                0 & 0 & 1 & 0 & 0 & 0 \\
                0 & 0 & 0 & 0 & 0 & 1 \\
                0 & \mathbf{H}_{31} & 0 & \mathbf{H}_{33} & \mathbf{H}_{34} & 0\\
            \end{bmatrix}
    \end{split}
\end{equation}
where
\begin{equation} \label{H_detail}
    \begin{split}
        \mathbf{H}_{31} &= nc((A_{sc} + A_{bs})(2\dot{\xi}^2 + \dot{z}^2)/V_t + \dot{\xi}\dot{z}(A_{bc} - A_{ss})/V_t )\\
        \mathbf{H}_{33} &= nc((A_{sc} + A_{bs})(\dot{\xi}+ \dot{z})/V_t + \dot{\xi}^2(A_{ss} - A_{bc})/V_t)\\
        \mathbf{H}_{34} &= l(T_1 \cos{(\beta + \theta)}  + T_2\cos{(\beta-\theta)} + cd \dot{\xi} V_t(A_{ss} - A_{bc}))
    \end{split}
\end{equation}
The symbols in  \eqref{J_detail} and \eqref{H_detail} are defined as follows
\begin{equation}
    \begin{split}
        cd &= 0.5 \rho C_D \\ 
        l &= \frac{1}{m_f+m_b} \\
        nc &= -0.5 \rho l C_D  \\
        \omega_{cm} &= \frac{\omega_b}{2} \\
        \alpha &= \tan^{-1}(\dot{z}/\dot{\xi}) \\
        V_t    &= \sqrt{\dot{\xi}^2 + \dot{z}^2}\\
        A_s     &= (h_b d_b)+(h_f d_f) \\
        A_b     &= (\omega_b d_b)+(\omega_f d_f)\\
        A_{sc} &= A_s \cos(\theta - \alpha) \\
        A_{ss} &= A_s \sin(\theta - \alpha) \\
        A_{bc} &= A_b \cos(\theta - \alpha) \\
        A_{bs} &= A_b \sin(\theta - \alpha)
    \end{split}
\end{equation}
All the basic constants value are written here
\begin{equation}
    \begin{split}
        \rho &= \SI{0.02}{\kilo\gram\per\metre\cubed}\\
        g &= \SI{3.711}{\meter\per\second\squared} \\
        \beta &= \SI[quotient-mode=fraction]{\pi/4}{\radian} \\
        C_D &= 0.2 \\
        m_f &= \SI{390}{\kilo\gram} \\
        \omega_f &= \SI{1}{\metre} \\
        h_f &= \SI{0.5}{\metre} \\
        d_f &= \SI{1}{\metre} \\
        m_b &= \SI{1510}{\kilo\gram} \\
        \omega_b &= \SI{3.2}{\metre} \\
        h_b &= \SI{2.5}{\metre} \\
        d_b &= \SI{2.9}{\metre} \\
        h_{cm} & = \SI{0.9421}{\metre} \\
    \end{split}
\end{equation}
Mapping between 3 dimensional process noise and 6 dimensional measurement noise $\mathbf{\Gamma}_k$ is 
\begin{equation}
    \mathbf{\Gamma}_k = 
    \begin{bmatrix}
        0 & 0 & 0 \\
        1 & 0 & 0 \\
        0 & 0 & 0 \\
        0 & 1 & 0 \\
        0 & 0 & 0 \\
        0 & 0 & 1 \\
    \end{bmatrix} 
\end{equation}
The fixed measurement noise is 
\begin{equation}
    \widetilde{\mathbf{v}}(t) = (1.0, 0.5, 0.025, 0.0225)^T
\end{equation}
The measurement noise covariance is set as 
\begin{equation}
    \R{k} = 
    \begin{bmatrix}
        1.0 & 0 & 0   & 0\\
        0 & 0.5 & 0   & 0 \\
        0 & 0 & 0.025 & 0 \\
        0 & 0 & 0     & 0.0025 \\
    \end{bmatrix}
\end{equation}

\subsection{Feedback Law of LQR controller}
\indent We need linearize the motion model in order to use LQR controller, e.g. calculating the Jacobian of motion model. We have calculated the Jacobian of motion model with respect to state $\mathbf{x}$. We also need the Jacobian with respect to control $\mathbf{u}$ and noise $\mathbf{w}$ respectively. The Jacobian with respect to control input is 
\begin{equation} \label{jac_control}
    \begin{split}
        \mathbf{U}(t) = 
            \begin{bmatrix}
                0 & 0 \\
                \sin(\theta + \beta)l & \sin(\theta -\beta)l \\
                0 & 0 \\
                \cos(\theta + \beta)l & \cos(\theta - \beta)l\\
                0 & 0 \\
                \frac{1}{I_\eta}(0.5\cos(\beta)\omega_b - \sin(\beta)h_{cm}) & -\mathbf{U}_{50}
            \end{bmatrix}
    \end{split}
\end{equation}
The Jacobian with respect to the noise is
\begin{equation}\label{jac_pnoise}
    \mathbf{W}(t) = 
        \begin{bmatrix}
            0 & 0 & 0 \\
            1 & 0 & 0 \\
            0 & 0 & 0 \\
            0 & 1 & 0 \\
            0 & 0 & 0 \\
            0 & 0 & 1 
        \end{bmatrix}
\end{equation}
Substitute $\x{ref}$ , $\mathbf{u}_{nom}$ into equation \ref{jac_pnoise} and \ref{jac_control} we can get the linearized value of jacobian at the desired point $\mathbf{F}_{\x{ref}}$ and $\mathbf{U}_{\mathbf{u}_{nom}}$. Then the linearized state space model around the desired state can be written as 
\begin{equation}
    \begin{split}
        \dot{\mathbf{x}} &= \mathbf{F}_{\x{ref}} \mathbf{x} +  \mathbf{U}_{\mathbf{u}_{nom}} \mathbf{u} \\
        \mathbf{y} &= \mathbf{x}
    \end{split}
\end{equation}
Then we can get the optimal gain matrix $K_{lin}$ by 
\begin{equation}
    \mathbf{K}_{lin} = \mathbf{R}_{con}^{-1}\mathbf{U}_{\mathbf{u}_{nom}}^T\mathbf{S}_{con}
\end{equation}
where $\mathbf{S}_{con}$ is the solution of of the associated Riccati equation
\begin{equation}
    \begin{split}
        & \mathbf{F}_{\x{ref}}^T \mathbf{S}_{con} + \mathbf{S}_{con}\mathbf{F_{\x{ref}}} - \mathbf{S}_{con}\mathbf{U}_{\mathbf{u}_{nom}}\mathbf{R}_{con}^{-1}\mathbf{U}_{\mathbf{u}_{nom}}^T \\ + & \mathbf{Q}_{con} = \mathbf{0}
    \end{split}
\end{equation}
the $R_{con}$ is a 2 by 2 diagonal matrix with its diagonal element (0.01, 0.01) and the $Q_{con}$ is a 6 by 6 diagonal matrix with its diagonal elements (200 15 200 15 10000 15). Finally we get our $\mathbf{K}_{lin}$
\begin{equation}
    \mathbf{K}_{lin} = 
        \begin{bmatrix}
            100.0 & -100.0 \\
            406.575 & -406.575 \\
            100.0 & 100.0 \\
            519.086 & 519.086 \\
            3053.285 & -3053.285 \\
            3140.470 & -3140.470 \\
        \end{bmatrix}^T
\end{equation}





\ifCLASSOPTIONcaptionsoff
  \newpage
\fi

\end{document}